%% file: Vector-auxiliary.tex
\documentclass[11pt]{article}
\usepackage{pdfpages}
\usepackage{geometry}  
\usepackage{chngcntr}
\counterwithin*{equation}{section}  
\usepackage[title,titletoc,toc]{appendix}
 \usepackage{hyperref}  
  \usepackage{color}
 \usepackage[most]{tcolorbox}
 \tcbset{colback=yellow!10!white, colframe=red!50!black, 
        highlight math style= {enhanced, 
            colframe=red,colback=red!10!white,boxsep=0pt}
        }        
\usepackage{graphicx}
\usepackage{amssymb}
\usepackage{amsmath}
\usepackage{amsfonts}
\usepackage{epstopdf}
\usepackage[export]{adjustbox}
\usepackage[compat=1.1.0]{tikz-feynman}
\newcommand{\olra}{\overleftrightarrow}
\newcommand{\p}{\partial}
\newcommand{\Dt}{\frac{D}{2}}
\newcommand{\dd}{\delta}
\newcommand{\lm}{\Lambda}
\newcommand{\al}{\alpha}
\newcommand{\lo}{\Lambda_0}
\newcommand{\hf}{\frac{1}{2}}
\newcommand{\be}{\begin{equation}}
\newcommand{\br}{\begin{eqnarray}}
\newcommand{\er}{\end{eqnarray}}
\newcommand{\ee}{\end{equation}}
\newcommand{\bt}{\begin{tabular}}
\newcommand{\et}{\end{tabular}}
\newcommand{\la}{\lambda}
\newcommand{\si}{\sigma}
\newcommand{\bp}{\bar p}

\newcommand{\eps}{\epsilon}
\newcommand{\CD}{{\cal D}}
\newcommand{\DD}{\Delta}
\newcommand{\ddz}{\frac{\partial}{\partial z}}
\newcommand{\ddx}{\frac{d}{dx}}
\newcommand{\Dp}{\frac{d^Dp}{(2\pi)^D}}

\newcommand{\fpi}{(4\pi)^\Dt}
\newcommand{\I}{\href{https://arxiv.org/abs/1706.03371}{I}}

\title{Bulk Gauge Fields and Holographic RG from Exact RG}
\author{Pavan Dharanipragada, Semanti Dutta and B. Sathiapalan\\Institute of Mathematical Sciences\\
CIT Campus, Tharamani\\ Chennai 600113, India\\
and\\ Homi Bhabha National Institute\\Training School Complex, Anushakti Nagar\\
Mumbai 400085, India\\bala@imsc.res.in}
\begin{document}

\maketitle
\begin{abstract}
Recently, a method was described for deriving  Holographic RG equation in $AdS_{D+1}$ space starting from an Exact RG equation of a $D$-dimensional boundary CFT \cite{Sathiapalan:2017}. The evolution operator corresponding to the Exact RG equation was rewritten as a functional integral of a $D+1$ dimensional field theory in $AdS_{D+1}$ space. This method has since been applied to elementary scalars and composite scalars in the $O(N)$ model \cite{Sathiapalan:2020}. In this paper, we apply this technique to the conserved vector current and the energy momentum tensor of a boundary CFT, the  $O(N)$ model at a fixed point. These composite spin one and spin two operators are represented by auxiliary fields and extend into the bulk as gauge fields and metric perturbations. We obtain, at the free level, the (gauge fixed) Maxwell  and Einstein actions. While the steps involved are motivated by the AdS/CFT correspondence, none of the steps logically require the AdS/CFT conjecture for their justification.
\end{abstract}
\newpage

\tableofcontents
\newpage

\section{Introduction}
 Holography has been appreciated for some time now as  a property of  any theory of gravity \cite{tHooft:1993dmi,Susskind:1994vu}. This has been realized in a fairly concrete way in the form of the 
 AdS/CFT correspondence, for which there is a lot of evidence by now \cite{Maldacena,Polyakov,Witten1,Witten2}\footnote{See \cite{Penedones2016} for a review and references.}. One of the interesting characteristics of this correspondence is that the radial coordinate of the bulk AdS can be thought of as the scale of the boundary theory. This correspondence then also implies 
that radial evolution in the bulk is to be interpreted as an RG evolution for the boundary theory [\cite{Akhmedov}-\cite{deHaro:2000}]. This has been termed ``Holographic RG".

In  \cite{Sathiapalan:2017}, (hereafter \I), it was shown that there is a way to obtain Holographic RG  starting from the Exact Renormalization Group (ERG) \cite{Wilson,Wegner,Wilson2,Polchinski} of the boundary theory. (For reviews of  ERG formalism and later developments, one may consult \cite{Wetterich,MorrisERG,Bagnuls1,Bagnuls2,Igarashi,Rosten:2010}.) In \I, the Polchinski ERG equation \cite{Polchinski} for a boundary $D$-dimensional scalar field theory is the starting point. The evolution operator correponding to this functional differential equation can be written as a functional integral with a $D+1$ dimensional free scalar field action. The action has a non-standard kinetic term. A field redefinition then renders this  action into a standard form of a free theory, but now in $AdS_{D+1}$ space. $AdS$ space is natural in this situation: in a CFT, dimensionless variables are obtained by compensating powers of the cutoff scale $\lm$. One obtains the same result in $AdS$ space where powers of the  radial coordinate $z$ are used because of the metric form
\be
ds^2 = \frac{dz^2 + d\vec x.d\vec x}{z^2}.
\ee
Thus, $\lm$ maps naturally to $1/z$. 

Therefore, in this approach the $d+1$ dimensional AdS ``bulk" theory functional integral is an ERG evolution operator for the ``boundary" theory action---as advocated in ``Holographic RG", but now without invoking the AdS/CFT conjecture for its justification.

In \I, the boundary theory was a free (elementary) scalar field theory, and only two point correlator was studied.
If this connection can be made more generally, this would give a new insight into the origin of holography, and place in a more central position the idea that the extra dimension of the higher dimensional theory is indeed the ``scale" of the lower dimensional theory. This is also suggestive of the possibility  that the AdS/CFT correspondence can be derived from Renormalization Group concepts \cite{SSLee:2010,SSLee:2012,Meloa:2019}. 

Towards this end, the techniques employed in \I$ $ 
 have been applied in other situations. In \cite{Sathiapalan:2019}, the case of the free elementary scalar with anomalous, (i.e, non canonical), dimension was discussed. In \cite{Sathiapalan:2020}, the $D$-dimensional O(N) model with $2<D<4$ was considered\footnote{A nice review of this model is \cite{Zinn-Justin,Moshe:2003}. Some extensions of these models have been studied---for e.g., \cite{Fei:2014} and more recently \cite{Jack:2021}. The holographic description has been studied in many papers. For eg \cite{Petkou:2003}-\cite{Sezgin:2002}]} The new ingredient here was that the RG evolution was for an action that describes the correlations of a {\em composite} scalar operator. The composite operator was described by using an auxiliary field. The ERG equation that describes the evolution of an action for this auxiliary field is different from the Polchinski ERG equation, (for an elementary field), in that it has additional terms.  The higher dimensional field theory in AdS space that determines the evolution operator for this modified ERG equation has the usual kinetic term (in AdS space). But the additional terms of the ERG equation give rise to interaction terms in the AdS field theory. In \cite{Sathiapalan:2020}, the cubic term was written down and shown to reproduce the 3-point correlation functions following the usual AdS/CFT prescription. This is not surprising: what is being done is nothing but Exact RG evolution---it is then guaranteed to give the correct correlation functions. What is not guaranteed and needs to be established are properties such as locality of the bulk theory. We do not have anything to say about this in this paper. In \cite{El-Showk:2011} general conditions on CFT's are postulated for the existence of a bulk space time dual and as pointed out there, the $O(N)$ models discussed in the present paper satisfies all the constraints. One of the requirements listed there is the existence of a large number of degrees of freedom in the CFT. Indeed this is confirmed in the present paper and also in \cite{Sathiapalan:2020} since large $N$ is used in an essential way. We have not used the other constraints in any obvious way in this paper - even  though the $O(N)$ models satisfy them.
Since the ERG approach presented in the present paper (and in \cite{Sathiapalan:2017,Sathiapalan:2019,Sathiapalan:2020}) is very general it may be interesting to see what role the other conditions mentioned in \cite{El-Showk:2011} play when a bulk dual is derived from first principles from ERG for other boundary theories. In particular the question of bulk locality mentioned above may play some role in this.
 
 The bulk dual of the $O(N)$ model is expected to be a higher spin theory \cite{Klebanov:2002} of the type described in \cite{Vasiliev:2003, Vasiliev:2004}. This connection has been investigated in \cite{Petkou:2003}-\cite{Sezgin:2002}]. In \cite{Jin:2015}, following a proposal in \cite{Douglas:2010}, higher spin equations have been derived using Polchinski's RG equations for a non local bilinear field that decribes an infinite tower of composite fields. This is similar in spirit to the approach in the present paper. In \cite{Jin:2015} some canonical tranformations in phase space are constructed, order by order, to make the linearized higher spin equations local. This may be  the counterpart of a field redefinition performed here that maps  the kinetic term of the ERG evolution operator exactly to a standard kinetic term in AdS. It would be interesting to understand this better.

One of the remarkable features of the AdS/CFT correspondence is that the bulk theory has gravity---the metric field being dual to the energy momentum tensor operator of the boundary theory. It also has gauge fields dual to conserved currents generating global symmetries of the boundary theory. In this paper, this is the main object of study. We derive ERG equations for the action that describes correlators of the energy momentum tensor and O(N) currents of the boundary theory. Once again, auxiliary fields are defined for the composite operators, and ERG equations are written for the action for these auxiliary fields. To our knowledge, this is not usually done in the treatment of composites in ERG, (see, for e.g., \cite{Igarashi}), but we find it very convenient for comparison with AdS/CFT results. We believe these techniques should be useful in the treatment of composite operators in other applications of ERG as well.  

The term ``auxiliary field" is used in this paper to stand for any dynamical field that does not have a tree level kinetic term, i.e., its classical equation of motion is algebraic. The word dynamical signifies then that, in the functional integral, this field is integrated over. This is in contrast with external fields that are non-dynamical, i.e., not integrated over.
Auxiliary fields have been used in physics for a long time in different situations. They have been used in the``Hubbard Stratonovich" transformation, as Lagrange multipliers, as gauge fields in Coset space sigma models, and $CP^N$ models\footnote{Although they start off without a kinetic term, quantum corrections typically generate a kinetic term as happens in the two dimensional $CP^N$ models for instance \cite{DAdda:1978,Witten:1978bc}.}. As mentioned above, recently they were used in \cite{Sathiapalan:2020} to obtain an ERG equation for an action decribing a composite scalar in the $O(N)$ model. The reason they are useful is the following. Composite operators in ERG are solutions to the {\em linearized} ERG equation\cite{Igarashi}. They are useful for calculating correlation with elementary fields using the low energy Wilson Action. However, they are not useful if multiple composite operators are present in the correlator because that makes it {\em non-linear}. But, this is precisely the kind of objects that are of interest in CFT's. Auxiliary fields turn out to be useful in this situation. It is possible to write down actions for them just as for elementary scalars.
One can also introduce sources for composite operator and calculate the generating functional. It will be seen below that the action for auxiliary fields is  the (logarithm of the) Fourier transform of the generating functional.

In the original version of the AdS/CFT correspondence (and also in \cite{Sathiapalan:2020}), sources for composite operators become dynamical fields in the bulk theory. This gives the generating function. In this paper, it is an auxiliary field of the boundary theory that extend into the bulk as a dynamical field. This gives the Fourier transform of the generating function, which is actually the Wilson action for the auxiliary field.

In this paper, we apply these techniques to the spin 1 vector and the spin 2 tensor operators. The vector is a conserved current for the $O(N)$ symmetry, and the spin 2 operator is the energy momentum tensor. We show in this paper that on mapping to AdS using the techniques of \cite{Sathiapalan:2017,Sathiapalan:2020}, they become bulk Yang-Mills gauge fields and the bulk metric tensor, i.e., gravitational fluctuations. More precisely, we find that if $a_\mu(z,x)$ is the bulk gauge field, and $J_\mu(x)$ is the boundary current, then as $z\to 0$,

\be a_\mu(z,x) \approx z^{D-2} J_\mu(x),\ee and for the bulk metric perturbation $h_{\mu\nu}(x)$, and boundary energy momentum tensor $T_{\mu\nu}$,
\be
h_{\mu\nu}(z,x)\approx z^{D-2} T_{\mu\nu}(x).
\ee

Furthermore, the bulk field theory dictates the Exact RG evolution of the boundary (Wilson) action for the auxiliary fields that stand for the current and energy momentum tensor respectively\footnote{It is also possible to write ERG equations for the boundary  Generating Functional but we do not do so in this paper.}.

We concentrate in this paper on the free (bulk) theory.
That is to say, we derive the quadratic kinetic term in AdS space for the gauge field and the graviton. The leading term, i.e., the kinetic term, is the Maxwell action and the linearized Einstein-Hilbert action respectively. 

The issue of maintaining gauge invariance in the presence of a cutoff does not arise at this stage because we make a gauge choice. The Ward Identity has then to be imposed when we consider the effect of interactions. But, we do not consider interactions in this paper, and we are primarily interested in a ``proof of concept" of the idea  that dynamical gauge fields and dynamical gravity in AdS arise naturally out of ERG. 
Dealing with interactions involves the technical question of how to maintain gauge invariance in ERG in the presence of a finite cutoff \cite{Igarashi},[\cite{Becchi}-\cite{Sonoda:2007dj},\cite{Morris:1999}-\cite{Arnone:2000q}], or, in the case of gravity, diffeomorphism invariance. This is left for the future.

This paper is organized as follows.
 In Section \ref{scalar}, we give some background on auxiliary field. We start with the free scalar O(N) theory. The auxiliary field standing for the $\phi^2$ operator has self interactions and is described by a non trivial Wilson action. The interacting theory is then considered and the calculation of anomalous dimension of $\phi^2$ is done in terms of auxiliary fields.
 
 In Section \ref{ERG}, we study the Wilson action for the auxiliary field for $\phi^2$ in the free scalar theory and write down the ERG equation satisfied by it. This technique for obtaining an ERG equation for a composite operator underlies the rest of the results in this paper---which apply this idea to other composites with spin. 
 
 In Section \ref{vector}, we discuss the vector current for the O(N) model at the Wilson Fisher fixed point and introduce an auxiliary field for this.  The same construction works for both the free and interacting fixed point theories. 
 
 In Section \ref{polch}, an ERG equation is obtained for the vector action. The evolution operator is written as a $D+1$ dimensional field theory.
 
 In Section \ref{map}, we map this action to  AdS space and show that Maxwell's equation is obtained.
 
 In Section \ref{tensor}, we repeat the same steps for the energy momentum tensor of the free scalar theory, and give the mapping to AdS where the linearized Einstein equations are obtained.
 
 We conclude in Section \ref{con} with a summary and dicussion of results.

\input{Auxiliary-Field-for-Scalar-Composites_input.tex}
\input{Regulator-and-ERG_input.tex}
\input{Vector-Auxiliary-Fields_input.tex}

\input{Setting-up-the-ERG-equation_input.tex}
\input{ERGtensor_input.tex}
\input{Summary_input.tex}

\begin{appendices}
\input{AnomDimAppendix_input.tex}

\input{RegERGAppendix_input.tex}
\input{ERGVector_input.tex}
\input{f-constraints_input.tex}

\input{tensorAppendix_input.tex}
\input{tensor-compressed_input.tex}

\end{appendices}

\input{bibliography_input.tex}
\end{document}

%% file: Auxiliary-Field-for-Scalar-Composites_input.tex
\section{Background: Auxiliary Field for Scalar Composites}\label{scalar}

The purpose of this section is to illustrate the use of auxiliary fields in the $O(N)$ model. First we consider a simple theory with ordinary integral, then elevate the variable to field and have same results on free field theory. In both cases, we find the partition function. Next, we conisder an interacting field theory of quartic interaction. In this case, the calculation of anomalous dimension of $\phi^2$ operator using auxiliary field action is illustrated. Readers who are familiar with this can go directly to Section \ref{ERG}. 

\subsection{Toy Model}
To illustrate the technique of auxiliary fields used in this paper and in \cite{Sathiapalan:2020}, we take a simple model involving ordinary integrals where everything can be done explicitly.  Note that the ``source" $J$ is for the composite $y^2$ and furthermore is introduced with an $i$. Thus, formally it is a Fourier transform---as it would be in Minkowski space, though here we consider Euclidean space theories.

We introduce the variable $s$ that stands for the ``composite" $y^2$. An action $S[s]$ for $s$ can be used to compute correlations of $s$. The steps, while unnecessary for this simple problem, are such that they can be generalized to field theories \cite{Sathiapalan:2020}.

Consider 
\br
Z[J]&=&\int dy ~e^{-\hf a y^2 -\la y^4 +iJy^2}\nonumber \\
&=& \int ds~ \int dy~ \dd(s-y^2)e^{-\hf a y^2 -\la s^2 +iJs}\nonumber \\
 \nonumber &=&\int ds~e^{iJ s -\la s^2} \int \frac{dx}{\sqrt {2\pi}}~e^{ixs -\hf \ln (a+2ix)}\\
 &=& \int ds~e^{iJs -S[s,\la]},
\er
where 
\br
\nonumber e^{-S[s,\la]}&=& \int \frac{dx}{\sqrt {2\pi}}~~e^{ixs -\hf \ln (a+2ix)-\la s^2}\\
&=&\int \frac{dJ}{2\pi}e^{-iJs}Z[J].
\er

\begin{itemize}
\item {\bf Case 1:~~$\boldsymbol {\la =0}$}
We can proceed in two ways.
\begin{enumerate}
\item 
Do the $s$ integral first:
\[
Z[J]=\int  dx~ \sqrt{2\pi} \dd(x+J)e^{-\hf \ln (a+2ix)}= \sqrt{2\pi} e^{-\hf \ln (a-2iJ)}.
\]

\item Do the $x$ integral first:
\[
Z[J]=\int ds e^{-S[s]+iJs},
\]
where
\be  \label{2}
e^{-S[s]} =\int \frac{dx}{\sqrt{2\pi}}e^{isx -\hf \ln (a+2ix)}
\ee
\[
=\int \frac{dx}{\sqrt{2\pi}}e^{isx}\frac{1}{\sqrt{a+2ix}}.
\]
This can be done using contour integrals (or by consulting the Bateman manuscript project).
\br 
e^{-S[s]}&=& e^{-\hf as}s^{-\hf},~~~s>0\nonumber\\
&=& 0,~~~~s<0.
\er

\end{enumerate}

\item {\bf Case 2:~~$\boldsymbol {\la =\infty}$}

\be
Z[J]=\int ds~\int \frac{dx}{2\pi}~ e^{ixs+iJs -\la s^2 -\hf \ln (a+2ix)}.
\ee
Do the $s$ integral. Then,
\[
Z[J]= \int \frac{dx}{\sqrt{2\pi}}e^{-\hf \frac{(x+J)^2}{2\la}-\hf \ln (a+2ix)}.
\] 
Now we take $\la \to \infty$, but define 
$J'= \frac {J}{2\la}$ as finite.
\[
Z[J']= e^{-\frac{J'^2 \la}{4}}  \int \frac{dx}{\sqrt{2\pi}} e^{-J'x -\hf \ln (a+2ix)}.
\]
We define 
\be   \label{3}
Z'[J']= \int \frac{dx}{\sqrt{2\pi}} e^{-J'x -\hf \ln (a+2ix)}.
\ee

\end{itemize}
\vspace{0.2 in}

Now compare \eqref{3} with \eqref{2}.  We see that

\begin{center}
\boxed{
Z'[J'] (\la =\infty)= e^{-S[iJ']}(\la=0)
}.
\end{center}

This is the toy model version of the duality between the two fixed points of the $O(N)$ model \cite{Klebanov:2002}. 

\subsection{Free Field Theory} 

Consider free field theory:
\begin{align}
	Z[J]&=\int {\cal D}\phi~e^{-S_B[\phi,J]}\nonumber\\
	&=\int {\cal D}\phi e^{-\int \hf \phi \DD^{-1}\phi + i\int J\phi^2}.
\end{align}

It is best to think of $J\phi^2$ as an interaction with an external $J$. We have a factor of $i$ as would have been there in Minkowski space. Since $J$ is an external field we are free to do this even in Euclidean space. In Euclidean space the final correlation functions are real, but, at intermediate stages, complex objects will be encountered. There is nothing unphysical about this.
Thus let us rewrite using a Lagrange multiplier field $\chi$ that imposes a delta function constraint as

\begin{align}
Z[J] =& \int {\cal D}\si \int {\cal D}\chi \int {\cal D}\phi~ e^{i\int \chi(\si-\phi^2)} e^{-\int \hf \phi \DD^{-1}\phi + iJ\si} \label{zj}\\
\label{10}
=& \int {\cal D}\si \int {\cal D}\chi~
e^{i\int \chi \si - \hf Tr \ln(\DD^{-1}+ 2i\chi) +i\int J \si}\\
\label{11}
=&
\int {\cal D}\si ~e^{-S[\si]+i\int J\si}.
\end{align}
This defines $S[\si]$ which can be used to calculate correlation functions of $\si$, which is the auxiliary field representing the composite $\phi^2$. If one does the $\si$ integral in \eqref{10}, one obtains a delta function, and subsequently,
\begin{align}
	Z[J]&= e^{-\hf Tr \ln (\DD^{-1}-2iJ)}\nonumber\\
	&\approx e^{-\hf Tr \ln (1-2i\DD J)}.
\end{align}

\subsection{Interacting Field Theory: Anomalous Dimension of Composite Operators Using Fundamental Fields and Auxiliary Fields}
\begin{figure}[ht] 
		\centering 
		\begin{tikzpicture}
			\begin{feynman}
				\vertex (i) {$\frac12m^2$}; 
				\vertex[right= 1.25cm of i] (a);
				\vertex[above right=0.75cm of a] (au);
				\vertex[below right=0.75cm of a] (ab);
				\vertex[right=2cm of a] (b);
				\vertex[above left=0.75cm of b] (bu);
				\vertex[below left=0.75cm of b] (bb);
				\vertex[right=0.25cm of b] (b1) {$\frac{\lambda}{4!}$};
				\vertex[right= 1cm of b, above right= 0.75 cm of b] (f1);
				\vertex[right= 1cm of b, below right= 0.75 cm of b] (f2);
				\diagram* {
					(i) -- [scalar](a)
					-- (au)
					--[scalar, quarter left](bu)
					-- (b)
					-- (bb)
					--[scalar, quarter left](ab)
					-- (a),
					(f1) -- (b),
					(f2)--(b)
				};
			\end{feynman}
		\end{tikzpicture}
	\caption{A sample one loop diagram contributing to the anomalous dimension of $\phi^2$}
	\label{fig:anomphi2}
	\end{figure}
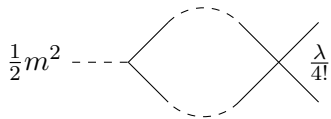

We now consider the interacting theory. We calculate anomalous dimensions of some operators using auxiliary fields. Let us consider the action
\be	\label{S1}
S=\int_x [ \hf \p_\mu \phi^I \p^\mu \phi^I +\frac{\la}{4!} (\phi^I\phi^I)(\phi^J\phi^J) + \hf m^2(x) \phi^I \phi^I].
\ee
The vertices are given in Figure \ref{fig:anomphi2}. 
$m^2(x)$ is being used as a source for the mass operator.

\subsubsection{Anomalous Dimension Using the Fundamental Fields}

The combinatorics of contracting legs in Figure \ref{fig:anomphi2} gives
\[
\sum_{A,I,J} (2 \dd^{AI}\dd^{AI}\phi^J\phi^J +2 \dd^{AJ}\dd^{AJ}\phi^I\phi^I +2.2.2 \dd^{AI}\dd^{AJ}\phi^J\phi^I)
\]
\be
=(4N +8)\phi^I\phi^I.
\ee
This finally gives a one loop contribution in $\phi^I \phi^I$,
\begin{align*}
	[\phi^I\phi^I](k)&=\frac{\la}{4!} (4N+8) \int \Dp  \frac{1}{p^2(p+k)^2}(\phi^I\phi^I)(k)\\
	&=
	\frac{(4N+8).}{4!} \frac{\la}{(4\pi)^2}\frac{1}{\eps}(\phi^I\phi^I)(k)\approx\frac{(4N+8)}{4!} \frac{\la}{(4\pi)^2}2t~(\phi^I\phi^I)(k),
\end{align*}
where we have replaced $\frac{1}{\eps}$ by $\ln \lm^2 \approx 2t$.
Thus,
\be  \label{gammaN}
\boxed{\frac{d}{dt}[\phi^I\phi^I]=\frac{(4N+8)}{4!} \frac{\la}{(4\pi)^2}2(\phi^I\phi^I)(k)}.
\ee
%

For large $N$ calculations, it is convenient to let $\frac{\la}{4!}= \frac{\bar \la}{2N}$.
So,
\be  \label{gamma}
\boxed{\gamma=\frac{4\bar \la}{(4\pi)^2}}.
\ee

\subsubsection{Using Auxiliary Fields}
\label{ad}
Consider the same action and repeat the steps in earlier sections:
\be
S=\int_x \bigg[ \hf \p_\mu \phi^I \p^\mu \phi^I +\frac{\bar \la}{2N} (\phi^I\phi^I)(\phi^J\phi^J) \bigg].
\ee
We now rescale $\phi^I \to N^{1/4}\phi^I$ so that the correlator $\langle\phi^I \phi^I(x) \phi^I \phi^I(0)\rangle \approx O(1)$.

\vspace{0.1 in}
Introduce a Lagrange multiplier $\chi$ to enforce $\sigma=\phi^I\phi^I$.
\[
S=\int_x [\sqrt N ( \hf \p_\mu \phi^I \p^\mu \phi^I )+\frac{\bar\la}{2} \sigma^2 + \chi(\sigma -\phi^I\phi^I) ].
\]
We will also add and subtract a terms $\chi_0 \sigma$ which can be written as:
\be  \label{SN}
S=\int_x  [\sqrt N( \hf \p_\mu \phi^I \p^\mu \phi^I +\hf m^2 \phi^I\phi^I)+\frac{\bar\la}{2} \sigma^2 + \chi(\sigma -\phi^I\phi^I) +\chi_0\sigma ].
\ee
And,
\be  \label{Z}
Z[J]=\int {\cal D}\chi {\cal D} \sigma {\cal D}\phi^Ie^{-S+\int _x J\sigma}.
\ee
$\chi_0$ is not physical for the moment. But we will choose it as vev for $\chi$ so that there is no linear term in $\chi$---``no tadpole condition"; (see below). This gives $\chi_0$ a physical significance.
Doing the $\phi^I$ integral gives:
\[
\int {\cal D}\phi^I e^ {-\hf\phi^I[\sqrt N(\Box +m^2)-2\chi]\phi^I}= e^{-\frac{N}{2}Tr \ln [1-\frac{2}{\sqrt N} \Delta \chi]},
\]
where field independent terms have been dropped. $(\Box +m^2)_x\Delta(x-y)=\dd(x-y)$ defines $\Delta$. One can also do the $\sigma$ integral to get
\be   \label{Z1}
Z[J]= \int {\cal D}\chi e^{-\hf N Tr \ln [1-2 \frac{\Delta\chi}{\sqrt N}]+\int \frac{(\chi+\chi_0-J)^2}{2\bar \la}}.
\ee
Expanding the log,
\be   \label{Z2}
\boxed{Z[J]=  \int {\cal D}\chi~e^{\frac{N}{2}Tr[\frac{2\Delta\chi}{\sqrt N} + \hf (\frac{2\Delta\chi}{\sqrt N})^2 +\frac{1}{3}(\frac{2\Delta\chi}{\sqrt N})^3+....] + \int \frac{(\chi-J)^2}{2\bar \la} +\int \frac{\chi \chi_0}{\bar \la} -\int \frac{J\chi_0}{\bar \la}+\int \frac{\chi_0^2}{2\bar \la}}}.
\ee
The last two terms are field independent and can be ignored.

\paragraph{Leading order}

\begin{itemize}
\item Now $\chi_0$ will be chosen so that the term linear in $\chi$ cancels:
\be
\boxed{\sqrt NTr [\Delta]=\frac{ \chi_0}{\bar \la}=-\sqrt N\frac{m^2}{2\bar \la}}.
\ee

\item Quadratic term in $\chi$:

\be   \label{S2}
S_2= -\bigg[Tr(\Delta \chi \Delta \chi) + \frac{(\chi-J)^2}{2\bar \la}\bigg].
\ee
Use
\[
\int dx e^{-\hf A(x-J)^2-\hf B x^2}= e^{-\hf J^2[\frac{AB}{A+B}]}=e^{-\frac{J^2}{2}[\frac{1}{A}+\frac{1}{B}]^{-1}}.
\]
Here $A= \frac{1}{\bar\la}$ and $B = 2 \Delta^2$.
So,
\be \label{lead_cor}
\langle\sigma \sigma\rangle=\Big[\bar \la + \frac{1}{2\Delta^2}\Big]^{-1}= 2\Delta^2\Big(\frac{1}{1+\bar \la 2\Delta^2}\Big)=-2\Delta^2(1-\bar \la 2 \Delta^2 +...).
\ee
\begin{figure}
\includegraphics[scale=0.3,center]{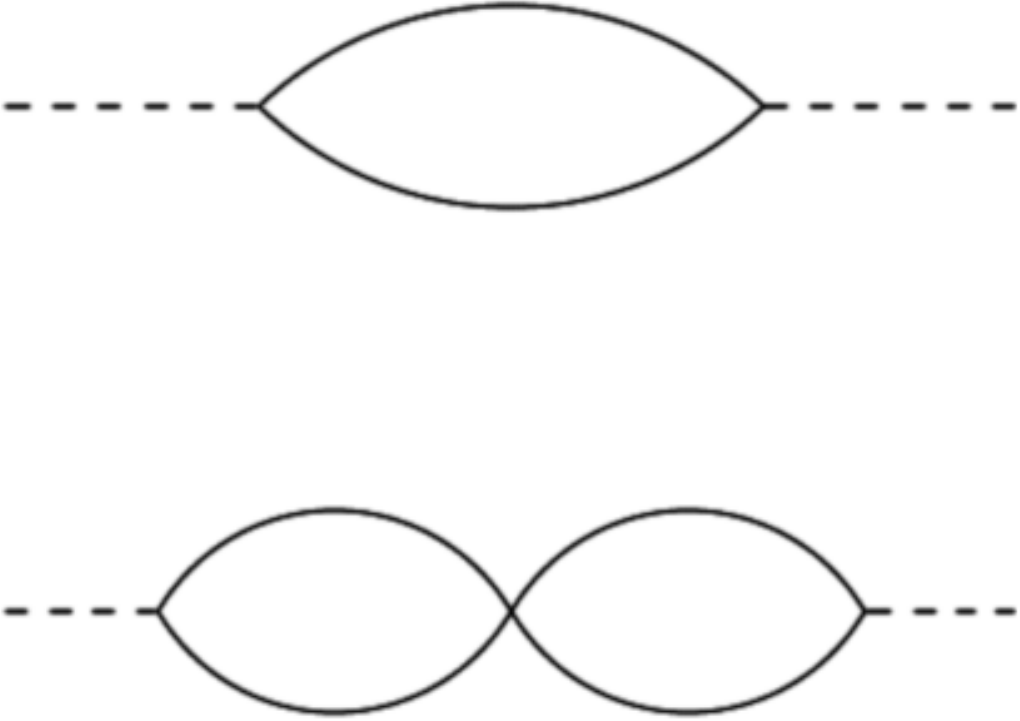}
\caption{Graphs contributing to anomalous dimension of $\si$}
\label{leading}
\end{figure}
The first two terms correspond to diagrams in Figure \ref{leading}.
This corresponds to an anomalous dimension, (deatils in Appendix \ref{anom}),
\be \label{lead_anom}
\gamma =  \frac{4\bar \la}{(4\pi)^2},
\ee
in agreement with \eqref{gamma}, but not in agreement with \eqref{gammaN}, which has a 1/N piece also.
\end{itemize}

Let us now turn to that correction.

\paragraph{Sub-leading 1/N correction}

We start with \eqref{SN} with appropriately rescaled fields:
\be
S=\int_x  [\sqrt N( \hf \p_\mu \phi^I \p^\mu \phi^I +\hf m^2 \phi^I\phi^I)+\frac{\bar\la}{2} \sigma^2 + \chi(\sigma -\phi^I\phi^I) +\chi_0\sigma ],
\ee
where we assume $\chi_0=-\hf \sqrt N m^2$. 

Then, as before,
\be  
Z[J]=\int {\cal D}\chi {\cal D} \sigma {\cal D}\phi^Ie^{-S+\int_x J\sigma}.
\ee
Doing the $\phi$ integral and then the $\sigma$ integral gives \eqref{Z1}:
\be   
Z[J]= \int {\cal D}\chi e^{-\hf N Tr \ln [1-2 \frac{\Delta\chi}{\sqrt N}]-\int \frac{(\chi+\chi_0+J)^2}{2\bar \la}}.
\ee
A field independent $ \hf N Tr\ln \sqrt N(\Box +m^2)$ has been dropped. $\DD=\frac{1}{\Box +m^2}$.

Note that the limit $\bar \la \to 0$ would set $\chi = -J-\chi_0$. We want the correction to this at $O(\la)$. As before $\chi_0$ will be chosen to cancel the tadpole in $\chi$. This gives a self consistent value for the vev of $\chi$. We can add a bare mass term $\hf r\phi^2$ for $\phi$ and use that as a tuning parameter corresponding to temperature. At the critical temperature $r_c$, $\chi_0=0$.  We are interested, as before, in the quadratic term in $\chi$ in order to obtain the anomalous dimension. So, for simplicity, we choose $\chi_0=0$.

We change variables: $\chi +J=\eta$. Then let $\eta = \sqrt {\bar \la} \eta'$.
We get
\be
Z[J]= \int {\cal D} \eta' e^{-\hf N Tr \ln [1+2 \frac{\DD J}{\sqrt N} - 2 \frac{\sqrt {\bar \la} \DD \eta'}{\sqrt N}]-\int \frac{\eta'^2}{2}}.\label{Zineta}
\ee

In the expnasion of the logarithm, the following two terms 
(cf. Figure \ref{leading}) give the anomalous dimension
\be
2{\bar \la} \int {\cal D} \eta' \{ Tr (\DD J \DD \eta') Tr (\DD J \DD \eta') + \frac{4 {\bar \la}}{N} Tr [(\DD J \DD \eta'\DD J \DD \eta')]\}e^{-\int \frac{\eta'^2}{2}}.\label{1/N}
\ee
Since the $\eta'$ propagator just gives a delta function in position space, the momentum loop integrals in both terms give the same value. The final answer thus has a factor
$2\frac{\bar \la }{N}(N+2)$. This is the correct $N$-dependence. Details are given in Appendix \ref{app:sublead}.


%% file: Regulator-and-ERG_input.tex
\section{Introducing a Regulator: Auxiliary Fields and ERG} \label{ERG}
In this section, we derive an Exact RG equation for the auxiliary field action. First, we show that the Wilson action for auxiliary field $\sigma$ is the natural object to find the correlators.  We then set up the ERG equation. We argue that it is necessary to perform an expansion in powers of $1/N$ in order to make contact with Holographic renormalization. Hence, at the end, we transfer all the results to N-scalar field theory. The ERG equation we derive has a linear term, and cubic and higher order terms at subleading order, in addition to the usual quadratic term present in Polchinski's equation. The linear term can be cancelled by redefinition of the auxiliary field. The cubic term is of the form derived in \cite{Sathiapalan:2020}.

\subsection{Regulated Theory}

We have been fairly cavalier about defining composite fields in the last section because it is a free theory. As a  preparation for the interacting theory let us rectify this by introducing a regulator so that there are no short distance singularities. We will then be able to define an ERG equation.

We start with
\be
Z[J]= \int {\cal D}\phi~ e^{-\hf \int \phi \DD^{-1} \phi -S_{B,I}[\phi,J]}.
\ee
We have included the dependence on the source $J$ as part of $S_{B,I}$ to be completely general. For the free theory, if we are interested in the composite $\phi^2$ we take
\be 
S_{B,I}[\phi,J]= -i\int J\phi^2 =-i\int J\si,
\ee
where we have anticipated that an auxiliary field $\si$ will be introduced to stand for $\phi^2$.

We write $\phi=\phi_l+\phi_h$ and $\DD=\DD_l+\DD_h$ as usual\footnote{See \cite{Igarashi} for a compact review of ERG formalism}, and write
\[
Z[J]=\int {\cal D}\si~ e^{ i\int J \si} \int {\cal D}\chi ~e^{i\int \chi\si}\int {\cal D}\phi_l~ e^{-\hf \int \phi_l \DD^{-1}_l \phi_l -i\chi \phi_l^2}\int {\cal D}\phi_h~ e^{-\hf \int \phi_h \DD^{-1}_h \phi_h -i\chi (\phi_h^2 + 2 \phi_h\phi_l)}.
\]
We take $\DD = \frac{K_0}{p^2}$ and $\DD_l= \frac{K}{p^2}$. Here $K_0$ has a UV cutoff at $\lo$ and $K$ at $\lm$. So, $\DD_l$ propagates frequencies below $\lm$, $\DD$ propagates frequencies below $\lo$ and $\DD_h$ propagates frequencies between $\lm$ and $\lo$. Note that this is for the field $\phi$. The auxiliary fields $\chi, \si$ have no such regulators.
 Nevertheless, since the only interactions these fields have are of the form $\chi \phi^2$, it is clear that modes of $\chi$ (and therefore $\si$) that have frequencies greater than $\approx 2\lo$ do not have any interactions, and so, play no role. Thus the theory is well defined in the UV. This is true even for the interacting theories that will be considered later. We now integrate out $\phi_h$, and define a Wilson action for $\phi_l$.
 
 We define first the interacting part of the Wilson action
\begin{align}
\label{2.13}
e^{-S_{\lm,I}[\phi_l,J]} &= \int {\cal D}\phi_h
~e^{-\hf\int \phi_h \DD^{-1}_h\phi_h - S_{B,I}[\phi_l+\phi_h,J]}\\
\label{2.14}
&\equiv\int {\cal D}\si e^{i\int J\si - S_{\lm,I}[\phi_l,\si]}.
\end{align}
For the case at hand, $S_{B,I}=-i\int J\phi^2 $. In the more general interacting case, there will be other terms. This defines $S_{\lm,I}[\phi_l,\si]$. This can be used to calculate $\si$-correlators---equivalently correlators of the composite object $\phi^2$.
 \be   \label{2.15}
e^{-S_{\lm,I}[\phi_l,\si]}= \int {\cal D}\chi ~
e^{i\int \chi\si}e^{-i\int \chi \phi_l^2}\int {\cal D}\phi_h~ e^{-\hf \int \phi_h \DD^{-1}_h \phi_h -i\chi (\phi_h^2 + 2 \phi_h\phi_l)},
\ee
and then the full Wilson action is
\be  \label{2.16}
S_\lm[\phi_l,J]= \hf \int \phi_l \DD^{-1}_l \phi_l+S_{\lm,I}[\phi_l,J],
\ee
and also
\[
S_\lm[\phi_l,\si]= \hf \int \phi_l \DD^{-1}_l \phi_l+S_{\lm,I}[\phi_l,\si],
\]
with
\be  \label{2.17}
e^{-S_\lm[\phi_l,J]}=\int {\cal D} \si ~e^{i\int J\si-S_\lm[\phi_l,\si]}.
\ee
Correlations of $\si$ fields are calculated as
\begin{center}
\boxed{
\langle \si (x_1)...\si(x_n)\rangle =
\int{\cal D}\si \int {\cal D}\phi_l ~\si (x_1)...\si(x_n)~e^{-S_\lm[\phi_l,\si]}
=
\langle \phi^2(x_1).... \phi^2(x_n)\rangle_B
}.

\end{center}

\paragraph{Note:} In ERG literature, the composite operator $[\phi^2]_\lm$ is defined as follows: if we start with $\phi^2$ in a bare theory and follow its linearized evolution, one obtains $[\phi^2]_\lm$. Thus, schematically, if $U$ denotes the evolution operator for ERG,
\[
[\phi^2]_\lm= U\phi^2 U^{-1},
\]
and 
\be 
\int {\cal D}\phi_l ~[\phi^2]_\lm e^{-S_\lm[\phi_l]} = \int {\cal D}\phi~ \phi^2e^{-S_B[\phi]},
\ee
and furthermore, in momentum space \cite{Igarashi},
\be 
\prod_{i=1}^n\frac{K_0(p_i)}{K(p_i)}\int {\cal D}\phi_l ~[\phi^2(p)]_\lm  \phi_l(p_1)...\phi_l(p_n) e^{-S_\lm[\phi_l]} = \int {\cal D}\phi~ \phi^2(p) \phi(p_1)...\phi(p_n)~e^{-S_B[\phi]}.
\ee

Thus, one might be tempted to equate $\si(x)$ with $[\phi^2]_\lm$ all along the RG trajectory. However, 
\[
[\phi^2(x)\phi^2(y)]_\lm \neq [\phi^2(x)]_\lm[\phi^2(y)]_\lm.
\]
Thus,
\[
\int {\cal D} \phi~ \phi^2(x) \phi^2(y) e^{-S_B[\phi]}=\int {\cal D} \phi~ [\phi^2(x) \phi^2(y)]_\lm e^{-S_\lm[\phi]} \neq \int {\cal D} \phi~ [\phi^2(x)]_\lm [\phi^2(y)]_\lm e^{-S_\lm[\phi]},
\]
whereas
\[
\int {\cal D} \phi~ \phi^2(x) \phi^2(y) e^{-S_B[\phi]}=\int {\cal D}\phi \int {\cal D}\sigma ~\si(x) \si(y) e^{-S_\lm[\phi,\si]}.
\]
Therefore, $\si(x)$ works more generally than $[\phi^2(x)]_\lm$ and is the correct object to work with.

\vspace{0.4 in}

We can also define a simpler object, $S_\lm[0,\si]$, where $\phi_l$ has been set to zero. This can be used to calculate
\be
\langle \si (x_1)...\si(x_n)\rangle_\lm =
\int{\cal D}\si  ~\si (x_1)...\si(x_n)~e^{-S_\lm[0,\si]}.
\ee
The physical interpretation is that these are correlations in a theory with an IR cutoff at $\lm$. We can recover the correlation functions of the original theory by taking the limit $\lm \to 0$. 
\be
\lim_{\lm\to 0}\langle \si (x_1)...\si(x_n)\rangle_\lm =
\langle \si (x_1)...\si(x_n)\rangle_B
=\langle \phi^2(x_1)...\phi^2(x_n)\rangle _B.
\ee

We use the notation:
\be
S_\lm[\si]= S_\lm[0,\si],~~~S_\lm[J]=S_\lm[0,J].
\ee
From \eqref{2.13} and \eqref{2.16},
\be  \label{2.25}
Z_\lm[\phi_l,J]=e^{-S_\lm[\phi_l,J]}=e^{-\hf \int \phi_l\DD_l^{-1}\phi_l}\int {\cal D}\phi_h ~e^{-\hf \int \phi_h\DD^{-1}_h \phi_h +i\int J(\phi_l+\phi_h)^2 }.
\ee
This is the generating functional of a theory with IR cutoff $\lm$, and is also a Wilson action due to its dependence on $\phi_l$.

The full generating functional (independent of $\lm$) is  
\be
Z[J]= \int {\cal D}\phi_l ~e^{-S_\lm[\phi_l,J]}.
\ee
It is also true that
\be
\lim_{\lm \to 0} Z_\lm[0,J]=Z[J].
\ee
Thus, if we could determine $Z_\lm[0,J]$ for general $\lm$ we can evaluate $Z[J]$.
We are thus interested in obtaining an equation for $S_\lm[\si]$ or $S_\lm[J]$.

\subsection{ERG equation of $S_{\Lambda,I}[\phi_l,J]$}

$S_{\lm,I}[\phi_l,J]$ defined in \eqref{2.25} obeys Polchinski's equation. From this one can derive an equation for $S_{\lm,I}[\phi_l,\si]$ using \eqref{2.14}. Then setting $\phi_l=0$ in these equations gives equations for $S_\lm[\si]$ and $S_\lm[J]$.

Polchinski's ERG Equation is
\be  \label{polchERG}
\frac{\p}{\p t}e^{-S_{\lm,I}[\phi_l,J]}=\hf \int_x\int_y \dot\DD_{hxy}\frac{\dd^2}{\dd \phi_l(x)\dd \phi_l(y)}e^{-S_{\lm,I}[\phi_l,J]}.
\ee
$S_{\lm,I}[\phi_l,J]$ is given by doing the Gaussian integral in \eqref{2.25}, and is
\be  \label{2.28}
S_{\lm,I}[\phi_l,J]= -\int _x iJ(x)\phi_l^2(x)
+2\int_x\int_y J\phi_l(x)(\frac{\DD_h}{1-2i\DD_hJ})_{xy}J\phi_l(y) +\hf Tr\ln[1-2i\DD_h J].
\ee
Simplifying:
\be  \label{2.29}
S_{\lm,I}[\phi_l,J]=-\int_x\int_y J\phi_l(x)\Big(\frac{1}{1-2i\DD_hJ}\Big)_{xy} \phi_l(y) +\hf Tr\ln[1-2i\DD_h J].
\ee

In the limit $\phi_l=0$, the Polchinski's equation becomes, (see Appendix \ref{actionJ}),
\be \label{2.30}
\boxed{
\frac{\p}{\p t}e^{-S_{\lm,I}[\phi_l,J]}|_{\phi_l=0}= -\hf \int_x \int_y \left( \frac{2i \dot \Delta_h J}{1-2i\Delta_h J} \right)_{xy}
}.
\ee
It can also be shown that it is a fixed point. (See Appendix \ref{actionJ}).

\subsection{ERG equation for $S_{\lm}[\phi_l,\si]$}

We go back to \eqref{2.15} and do the $\phi_h$ integration to obtain
\begin{align}
e^{-S_{\lm,I}[\phi_l,\si]} &=
\int {\cal D}\chi ~e^{i\int \chi \si -i\int \chi \phi_l^2}e^{ -\hf Tr \ln [\frac{1 }{\DD_h}+2i\chi] }e^{-\hf \int_x\int_y(2\chi \phi_l(x))([\frac{1}{\DD_h}+2i\chi]^{-1})_{xy}(2\chi \phi_l(y))}\nonumber\\
&\label{2.38}
=\int {\cal D}\chi ~e^{i\int \chi \si -\hf Tr \ln [\frac{1}{\DD_h}+2i\chi]-i\int_x\int_y \chi \phi_l(x) ([1+2i\DD_h\chi]^{-1})_{xy}\phi_l(y) }\\
&\equiv\int {\cal D}\chi ~e^{i\int \chi \si-S_{\lm,I}[\phi_l,\chi]}.
\end{align}
Since the structure of $S_{\lm,I}[\phi_l,\chi]$ is exactly the same as that of $S_{\lm,I}[\phi_l,J]$, it is clear that it obeys Polchinski's ERG equation, and therefore, so does $S_{\lm,I}[\phi_l,\si]$, and is also a fixed point solution. 

Let us work out the ERG equation obeyed by $S_{\lm,I}[0,\si]$. This is essentially of the form in \eqref{2.30}:
\begin{align}
\frac{\p}{\p t}e^{-S_{\lm,I}[0,\si]}&=\hf\int_x\int_y\dot \DD_{hxy} \frac{\dd^2}{\dd \phi_l(x)\dd \phi_l(y)}e^{-S_{\lm,I}[\phi_l,\si]}|_{\phi_l=0}\nonumber\\
&= \hf\int {\cal D}\chi e^{i\int \chi \si}
\int_x\int_y \dot \DD_h(x-y)\bigg(-\frac {2i \chi}{1+2i \DD_h \chi}\bigg)_{xy}e^{-S_{\lm,I}[0,\chi]}\nonumber\\
\label{2.40}&
=
\hf\int_x\int_y \dot \DD_h(x-y)\bigg(-\frac {2 \frac{\dd}{\dd \si}}{1+2 \DD_h \frac{\dd }{\dd \si}}\bigg)_{xy} 
e^{-S_{\lm,I}[0,\si]}.
\end{align}

This a functional differential equation where only the quadratic term is of the Polchinski form. The linear term can be set to zero by a choice of counterterm linear in $\sigma$, i.e., a zero-momentum source term for $\si$. 
The cubic and higher terms were dealt with in \cite{Sathiapalan:2020} by using a $1/N$ approximation---which requires that we work with $N$ scalar fields.   
Order by order in $1/N$, one can replace the higher derivative terms by polynomials in $\si$. Thus, if the action is schematically of the form
\[
S[\si] \approx \hf\int \sigma \DD_h^{-2} \sigma + \frac{1}{\sqrt N} c_1 \int \sigma^3 + \frac{1}{N} c_2 \int \sigma^4+...,
\]
then a term $\frac{1}{\sqrt N} (\frac{\dd}{\dd \si})^3$ will contribute terms of the form 
$\frac{1}{\sqrt N}(\DD_h^{-2}\si)^3$ in the ERG equation. These can be interpreted as ``potential" terms in a Schrödinger-like differential equation. In fact, such terms constitute the potential terms in the evolution operator for this ERG equation.
These also become potential terms in the AdS action when one performs the mapping to AdS as described in \cite{Sathiapalan:2017}. This was worked out in \cite{Sathiapalan:2020} for the $O(N)$ model.

The conclusion is then that a small parameter $1/N$ needs to be introduced in order to make contact with Holographic RG. Let us proceed to do that by introducing $N$ scalar fields, $\phi^I$, and a normalization for the kinetic term that ensures that the propagator for $\si=\phi^I\phi^I$ is $O(1)$. 

\be
Z[J]~=\int {\cal D}\phi e^{-\sqrt N\int \hf \phi^I \DD^{-1}\phi^I + i\int J\phi^I\phi^I}.
\ee
This has the effect of replacing all $\DD_h$'s by $\frac{\DD_h}{\sqrt N}$ in the final equations.

\eqref{2.28} is modified to
\begin{align}
S_{\lm,I}[\phi_l^I,J]=& -\int _x iJ(x)\phi_l^I\phi_l^I(x)\\
&-2\int_x\int_y J\phi_l^I(x)(\frac{\DD_h/\sqrt N}{1-2i\DD_hJ/\sqrt N})_{xy}J\phi_l^I(y) +\frac{N}{2} Tr\ln[1-2i\DD_h J/\sqrt N],\nonumber
\end{align}
and \eqref{2.30} is modified to
\be  \label{2.43}
\hf\int_x\int_y {\dot\Delta_{hxy}\over \sqrt N}
\sum_I\frac{\dd^2S_{\lm,I}[\phi_l,J]}{\dd \phi_l^I(x)\dd \phi_l^I(y)}|_{\phi_l=0}=
-\frac{N}{2}\int_x\int_y (\frac {2i\dot \DD_h J/\sqrt N}{1-2i \DD_h J/\sqrt N})_{xy}.
\ee
\eqref{2.38} modified with appropriate powers of $N$ is
\begin{align}
e^{-S_{\lm,I}[\phi_l,\si]}&=
\int {\cal D}\chi ~e^{i\int \chi \si -i\int \chi \phi_l^2}e^{ -\frac{N}{2} Tr \ln [\frac{\sqrt N }{\DD_h}+2i\chi]^{-1} }e^{-\hf \int_x\int_y(2\chi \phi_l^I(x)([\frac{\sqrt N }{\DD_h}+2i\chi]^{-1})_{xy}2\chi \phi_l^I(y)}\nonumber\\
\label{2.51}
&=\int {\cal D}\chi ~e^{i\int \chi \si -\frac{N}{2} Tr \ln [\frac{\sqrt N }{\DD_h}+2i\chi]^{-1}-i\int_x\int_y \chi \phi_l^I(x) ([1+\frac{2i\DD_h\chi}{\sqrt N}]^{-1})_{xy}\phi_l^I(y) }.
\end{align}

We redefine $\si$ in order to cancel any tadpole contribution, and replace $\chi= \frac{1}{i} \frac{\delta}{\delta \si}$ at the leading order. (See Appendix \ref{actionsigma}.) The ERG equation is obtained as
\begin{align}
\frac{\p}{\p t}e^{-S_{\lm,I}[\si]}=&
\Big(
-\int_x\int_y \frac{d (\DD_h(x-y))^2}{dt}\frac{\dd^2}{\dd \si(x)\dd\si(y)}\\
&+ \frac{1}{i}\frac{4}{\sqrt N}\int_{x,y,z}\dot \DD_h(x-y)\DD_h(x-z)\DD_h(z-y)\frac{\dd^3}{\dd \si(x)\dd \si(y)\dd\si(z)}+....\Big)e^{-S_{\lm,I}[\si]}.\nonumber
\end{align}

This is the same form obtained in \cite{Sathiapalan:2020} for the ERG equation, but there it was at the Wilson-Fisher fixed point, and for the {\em generating functional} $W_\lm[J]$ rather than for the Wilson Action $S_\lm [\si]$. This is in accordance with the expected duality between the two fixed points of the $O(N)$ model.

This concludes our  discussion of auxiliary fields: how they can be used to describe composite operators, and how ERG equation in terms of auxiliary fields are obtained. We shall apply these ideas to vector and tensor composite operators in the coming sections.

%% file: Vector-Auxiliary-Fields_input.tex
\section{Vector Auxiliary Fields} \label{vector}

We consider the simplest conserved current in the boundary CFT---the current associated with the global O(N) symmetry. As before, we define an auxiliary vector field to stand in for the current. This field is defined as being dual to an external $O(N)$ gauge field. This ensures that current conservation is obeyed all along the RG trajectory as long as gauge invariance is. We show that the evolution operator corresponding to the ERG equation of the auxiliary field gives the kinetic term for an action of a theory of gauge boson in D+1 dimensions. As we set out to show the result only to leading order, we take the current to be effectively abelian in this paper.  This section deals with defining the auxiliary field and its action.   After  we have defined a suitable action for the auxiliary field, the ERG equation is worked out in the next section.  

\subsection{What action should we work with?}

A generalization of the auxiliary field technique that was used for scalar composites would lead us to the following action:
\be   
S_B=\int _x \Big[\hf  \sqrt N\p _\mu \phi^I \p^\mu \phi^I + \frac{\bar u}{2} \sigma^2 + \hf \sqrt N r \sigma + \chi(\sigma - \phi.\phi) - \chi_\mu^{AB}(\phi_A \olra{\p^\mu} \phi_B - \sigma^\mu_{AB})\Big].
\ee
Here, $\chi_\mu^{AB}$ is a Lagrange multiplier that imposes
\[
\si^\mu_{AB}= \phi_A\olra {\p^\mu} \phi_B.
\]
We need to improve on this because we want $\si _\mu^{AB}$ to always be the effective current---which may not be just $\phi^A \olra{\p_\mu} \phi^B$. So, we start by introducing an external gauge field $A^\mu_{IJ}$ to gauge the $O(N)$ symmetry. Thus we start with:
\be
S_B[\phi^I,\chi,\si,A_{\mu J}^I]=\int _x \Big[\hf  \sqrt ND _\mu \phi^I D^\mu \phi^I + \frac{\bar u}{2} \sigma^2 + \hf \sqrt N r \sigma + \chi(\sigma - \phi.\phi) \Big],
\ee
where 
\[
D_\mu \phi^I =\p_\mu \phi^I -A_{\mu J}^{I}\phi^J.
\]
For the $O(N)$ group, we don't have to distinguish between upstairs and downstairs indices, so, $A_{\mu I J}=A_\mu^{IJ}=A_{\mu J}^I$. The order from left to right is important; $A_{\mu J}^I=-A_{\mu I}^J$. 

\subsection{$S[\si _\mu,A^\mu]$ from $S_B$}

Define
\be
Z[A_\mu^{IJ}]= \int{\cal D}\phi^I e^{-S_B[\phi,A_\mu^{IJ}]}=\int{\cal D}\phi^I\int{\cal D}\sigma \int{\cal D}\chi~ e^{-S_B[\phi,\si,\chi,A_\mu^{IJ}]}.
\ee
After doing the $\si$ integral one obtains:
\be
Z[A_\mu^{IJ}]= \int{\cal D} \phi^I \int {\cal D}\chi e^{-S_B[\phi^I,\chi,A_\mu^{IJ}]},
\ee
with
\begin{align}
	S_B[\phi^I,\chi,A_\mu^{IJ}]=& \int _x \Big[\hf  \sqrt N \p _\mu \phi^I \p^\mu \phi^I +\sqrt NA_\mu^{IJ}(\phi^I \olra \p_\mu \phi^J)+\hf \sqrt N A_\mu^{IJ}A_\mu^{IK}\phi^J\phi^K\nonumber\\
	&
	+ \frac{(\chi+\hf \sqrt N r)^2}{2\bar u}  - \chi \phi.\phi \Big].
\end{align}

The $O(N)$ current is defined operationally as
\be
j_\mu^{IJ}= \frac{\dd}{\dd A^\mu_{IJ}}.
\ee
Thus,
\br
\langle j_\mu^{IJ}(x) \rangle &=& \frac{\dd}{\dd A^\mu_{IJ}(x)}Z[A],\\
\langle  j_\mu^{IJ}(x) j_\nu^{KL}(y) \rangle &=&\frac{\dd^2}{\dd A^\mu_{IJ}(x) \dd A^\nu_{KL}(y)}Z[A]. \label{delta}
\er

Let us introduce an auxiliary field $\si_\mu$ to stand for the current $J_\mu$
by means of a delta function:
\be
Z[A]=\int {\cal D}\si_\mu^{IJ} \dd\Big(\si_\mu ^{IJ}-\frac{\dd}{\dd A^\mu_{IJ}}\Big)Z[A],
\ee
and rewrite using Lagrange multiplier $\chi^\mu_{IJ}$:
\begin{align}
Z[A]&=\int {\cal D}\si_\mu^{IJ}\int {\cal D}\chi^\mu_{IJ}e^{i\int_x \chi^\mu_{IJ}(x)\big(\si_\mu^{IJ}(x)-\frac{\dd}{\dd A^\mu_{IJ}(x)}\big)}\int{\cal D}\phi^I\int {\cal D} \chi e^{-S_B[\phi^I,\chi,A_\mu^{IJ}]}\nonumber\\
&=
\int {\cal D}\si_\mu^{IJ}\int {\cal D}\chi^\mu_{IJ}e^{i\int_x \chi^\mu_{IJ}(x)\si_\mu^{IJ}(x)}\int{\cal D}\phi^I \int {\cal D} \chi e^{-S_B[\phi^I,\chi,(A^\mu_{IJ}-i\chi^\mu_{IJ})]}.
\end{align}

Note that $\chi^\mu_{AB}=-\chi^\mu_{BA}$ and has the same interaction as the external gauge field $A_\mu$. Thus the action has gauge invariance where we can transform $\chi^\mu$ instead of $A^\mu$, along with $\phi^I$, $\si_\mu^{IJ}$. One has to use the fact that the current is conserved, so that $\p^\mu\si_\mu^{IJ}=0$.

\vspace{0.1 in}

Shifting $i\chi_\mu \to i\chi_\mu+A_\mu$, we get,
\be
=\int {\cal D}\si_\mu^{IJ}e^{\int \si_\mu^{IJ}A^\mu_{IJ}}\int {\cal D}\chi^\mu_{IJ}e^{i\int_x \chi^\mu_{IJ}(x)\si_\mu^{IJ}(x)}\int{\cal D}\phi^I \int {\cal D} \chi e^{-S_B[\phi^I,\chi,(-i\chi^\mu_{IJ})]}.
\ee
This can be used to define the action for $\si_\mu^{IJ}$:
\be
Z[A_\mu]= \int {\cal D}\si_\mu^{IJ}e^{-S_B[\si_\mu^{IJ}] +\int \si_\mu^{IJ}A^\mu_{IJ}}.
\ee

One can evaluate $\sigma_\mu$ correlators with this action:
\begin{align}
\langle \si_\mu^{AB}(x_1) \si _\nu^{CD}(x_2)\rangle_B &\equiv
\int {\cal D}\si _\mu^{IJ}~~\si_\mu^{AB}(x_1) \si _\nu^{CD}(x_2)~e^{-S[\sigma_\mu^{AB}]}\\
&=
\int {\cal D}\si _\mu^{IJ} {\cal D}\phi^I {\cal D} \chi {\cal D} \chi^\mu_{AB}~\si_\mu^{AB}(x_1) \si _\nu^{CD}(x_2)e^{-S_B[\phi^I,\chi,-i\chi^\mu_{AB}]+i\int \chi^\mu_{AB}\sigma_\mu^{AB}}.\nonumber
\end{align}
Using \eqref{delta}, in terms of $\phi^I$,
\begin{align}
\langle \si_\mu^{AB}(x_1) \si _\nu^{CD}(x_2)\rangle_{B,A^\mu=0} =
&4N\langle (\phi^A \olra \p_\mu \phi^B)(\phi^C \olra \p_\nu \phi^D)\rangle\\
& +\sqrt N\langle(\dd ^{AC}\phi^B\phi^D -\dd ^{AD}\phi^B\phi^C+\dd ^{BD}\phi^A\phi^C-\dd ^{BC}\phi^A\phi^D)\rangle.\nonumber
\end{align}


\subsection{Gauge Invariance and Physical Degrees of Freedom}

Since $Z[A^\mu]$ is gauge invariant, one can choose the Lorentz gauge $\p_\mu A^\mu=0$. Thus we can define $A^{T\mu}$ that obeys $\p_\mu A^{T\mu}=0$.
$A^T_\mu$ can be defined by means of a projector
\be
A^T_\mu(q)= \Big(\dd_{\mu\nu}- \frac{q_\mu q_\nu}{q^2}\Big)A^\nu(q).
\ee
 Thus, our construction is modified as follows:
\begin{align*}
Z[A^{T}_{\mu}]&=\int {\cal D}\si_\mu^{T}\dd\bigg(\si_\mu^{T}-\frac{\dd}{\dd A^{T\mu}}\bigg)Z[A^{T}_{\mu}]\\
&=\int {\cal D}\si_\mu^{T}\int {\cal D}\chi^{T\mu} e^{\int \chi^{T\mu} \Big(\si_\mu^T-\frac{\dd}{\dd A^{T\mu}}\Big)}Z[A_\mu^T].
\end{align*}
Thus, we work entirely in a subspace that has the physical, transverse degrees of freedom. 

In the next section, we define a related object, $S_\lm[\si^T_\mu]$, that obeys an exact RG equation. This will be understood in physical terms as a current in a theory with an IR cutoff, $\lm$. 
This also removes the  $q \approx 0$ region where the projector may cause problems. The $\lm \to 0$ limit then gives $S[\si^T_\mu]$.

The strategy is to define the bulk action in terms of the transverse degrees of freedom. This can then be interpreted as a gauge fixed version of a gauge invariant bulk action in AdS space.

%% file: Setting-up-the-ERG-equation_input.tex
\section{ERG equation for Vector Action} \label{polch}

We now proceed to derive an ERG equation for the vector auxiliary field action. We impose transversality for the auxiliary fields. This is done by fixing  gauge for the external gauge field. The presence of a finite cutoff 
spoils manifest gauge invariance. These issues have been discussed in the literature \cite{Igarashi},[\cite{Becchi}-\cite{Sonoda:2007dj},\cite{Morris:1999}-\cite{Arnone:2000q}].The BRST method, described in \cite{Igarashi}, for instance, is adequate for our purpose. In fact, since we are only concerned about the kinetic term in this paper, we do not even have to worry about the BRST ghosts. So we will simply work in a fixed gauge. This means that the final equation and action for the bulk gauge field derived in Section \ref{map} are also obtained in a gauge fixed form.

These issues are also related to issues of renormalization of UV divergences faced in the discussions on holgraphic renormalization (see \cite{deHaro:2000} for instance).

\subsection{$S_\lm[\phi_l,\si_\mu]$ from  $S_\lm[\phi_l]$}
 
We now introduce momentum cutoffs $\lo,\lm$ in the kinetic term for $\phi$ and write down the expression for the partition function.
\be
e^{-S_{I,\lm}[\phi_l,A_\mu^T]}= \int {\cal D}\phi_h^I e^{-\hf \int \phi_h^I\DD_h^{-1}\phi_h^I -S_{B,I}[\phi_l+\phi_h,A_\mu^T]}.
\ee 

Note that $A_\mu$ has already been chosen to be transverse. We will drop the superscript $T$ from now on for simplicity.
Then,
\be
Z[A^\mu_{IJ}]=\int {\cal}D\phi_l e^{-S_\lm[\phi_l,A^\mu_{IJ}]}.
\ee
We can repeat the earlier steps to introduce $\si_\mu,\chi^\mu$, (where both are chosen to be transverse):
\begin{align}
Z[A]&=\int {\cal D}\si_\mu^{IJ}\int {\cal D}\chi^\mu_{IJ}e^{i\int_x \chi^\mu_{IJ}(x)\big(\si_\mu^{IJ}(x)-\frac{\dd}{\dd A^\mu_{IJ}(x)}\big)}\int{\cal D}\phi_l^I e^{-S_\lm[\phi_l,A_\mu^{IJ}]}\nonumber\\
&=
\int {\cal D}\si_\mu^{IJ}\int {\cal D}\chi^\mu_{IJ}e^{i\int_x \chi^\mu_{IJ}(x)\si_\mu^{IJ}(x)}\int{\cal D}\phi_l^I e^{-S_\lm[\phi_l^I,(A_\mu^{IJ}-i\chi^\mu_{IJ})]}.
\end{align}
Shifting variables as before,
\be
=\int {\cal D}\si_\mu^{IJ}e^{\int \si_\mu^{IJ}A^\mu_{IJ}}\int {\cal D}\chi^\mu_{IJ}e^{i\int_x \chi^\mu_{IJ}(x)\si_\mu^{IJ}(x)}\int{\cal D}\phi_l^I e^{-S_\lm[\phi_l^I,-i\chi^\mu_{IJ}]}.
\ee
Let us keep the $\phi_l$ integral separate and define
\be \label{za}
Z[A_\mu]\equiv\int {\cal D}\phi_l^I\int {\cal D}\si_\mu^{IJ}e^{-S_\lm[\phi_l^I,\si_\mu^{IJ}] +\int \si_\mu^{IJ}A^\mu_{IJ}}.
\ee
Clearly,
\be  \label{Ssigma}
e^{-S_\lm[\phi_l^I,\si_\mu^{IJ}]}=\int {\cal D}\chi^\mu_{IJ}e^{i\int_x \chi^\mu_{IJ}(x)\si_\mu^{IJ}(x)}
e^{-S_\lm[\phi_l^I,-i\chi^\mu_{IJ}]}.
\ee
$S_\lm[\phi_l^I,-i\chi^\mu_{IJ}]$ is a Wilson action for $\phi_l^I$ where $\chi^\mu_{IJ}$ is a fixed external field. If we remove from it $\hf\int \phi_l^I \DD_l^{-1}\phi_l^I$, we get $S_{I,\lm}[\phi_l^I,\chi^\mu_{IJ}]$, which obeys Polchinski's ERG equation. Using \eqref{Ssigma}, we see that $S_\lm[\phi_l,^I\si_\mu^{IJ}]$ also obeys the same equation. In the next subsection, we derive an explicit form for this equation.

It is important to note that just as we discussed in Section \ref{ERG}, $\si_\mu^{IJ}(x)$ {\em cannot} be understood as the composite current operator corresponding to the bare current $\phi^I \olra{\p_\mu} \phi^J(x)\equiv j_\mu^{IJ}(x)$, i.e.,
\[
\si_\mu^{IJ}(x)\neq [\phi^I \olra{\p_\mu} \phi^J(x)]_\lm~~(=[j_\mu^{IJ}(x)]_\lm).
\]
This is because
\[
\langle \si_\mu^{IJ}(x) \si_\nu^{KL}(y)\rangle = \langle [j_\mu^{IJ}(x)j_\nu^{KL}(y)]_\lm\rangle \neq \langle [j_\mu^{IJ}(x)]_\lm[j_\nu^{KL}(y)]_\lm \rangle.
\]

This means that if we are working with $S_\lm[\phi_l,A^\mu]$, we cannot use the composite operator $[j_\mu^{IJ}(x)]_\lm$ to calculate the correlator of currents. In particular, we should expect 
\[
\p_x^\mu \langle [j_\mu^{IJ}(x)]_\lm[j_\nu^{KL}(y)]_\lm \rangle  \neq 0
\]
even when $x\neq y$. The putative currrent-current correlator is not transverse, i.e., it does not give the real current-current correlator.
On the other hand, $\si_\mu$ can be used always for the current:
\[
\p_x^\mu \langle \si_\mu^{IJ}(x)\si_\nu^{KL}(y) \rangle = 0
\]
always when $x\neq y$.

\subsection{ERG equation for $S_\Lambda[\si_\mu^{IJ}]$}

Our starting point is the expression 
\be   \label{15}
e^{-S_\lm[\phi_l^I,\sigma_\mu^{IJ}]}=\int {\cal D}\phi_h^I {\cal D} \chi {\cal D} \chi_\mu^{AB}~e^{-S_B[\phi_l^I,\chi,\chi^\mu_{AB},\phi_h^I]+i\int\chi^\mu_{AB}\sigma_\mu^{AB}}.
\ee

And the usual Polchinski ERG equation will be written for the interacting part of $S_\lm[\phi_l^I,\si_\mu^{IJ}]$---defined as the part of the action with the kinetic term $\hf \phi_l^I\DD_l^{-1}\phi_l^I$ removed. 
Removing the kinetic term $\hf \phi_l^I\DD_l^{-1}\phi_l^I$ from both sides of \eqref{15}, we get
\be   \label{16}
e^{-S_{I,\lm}[\phi_l^I,\sigma_\mu^{IJ}]}=\int {\cal D}\phi_h^I {\cal D} \chi {\cal D} \chi^\mu_{AB}~e^{-S_{I,B}[\phi_l^I,\chi,\chi^\mu_{AB},\phi_h^I]+i\int \chi ^\mu_{AB}\sigma_\mu^{AB}}.
\ee

As mentioned in the last section,
the Wilson action $S_{\lm,I}[\phi_l,\si_\mu^{IJ}]$, in which $\si_\mu^{IJ}$ acts as an external source, obeys the Polchinski ERG equation, (derived in Appendix \ref{vectorapp}), (note the factor of $1/\sqrt N$  and also that $\dot \DD_h=-\dot \DD_l$):
 \be  \label{20}
\frac{\p }{\p t}e^{-S_{\lm,I}[\phi_l,\si_\mu^{IJ}]}=\hf \frac{1}{\sqrt N} \int_x\int_y \dot \DD _{hxy}\frac{\dd^2}{\dd \phi_l ^I(x)\dd \phi_l ^I(y)} e^{-S_{\lm,I}[\phi_l,\si_\mu^{IJ}]},
\ee 

 We are interested in the $\si_\mu^{IJ}$ dependence. So, after the functional differentiation has been performed, we can evaluate both sides at $\phi_l^I=0$, and get an equation for just the $\si _\mu^{IJ}$ dependence. While this simplifies the ERG equation, it means that $\lm$ is a physical IR cutoff in this theory, and {\bf only in the limit $\lm \to 0$ is the original physics of the infinite volume theory recovered.} Since the ERG equation determines $S_\lm[\si_\mu^{IJ}]$ for general $\lm$, this is not a limitation.

The details of deriving the ERG equation are given in Appendix \ref{vectorapp}. We state salient points below. 

\begin{enumerate}
\item
 The ERG equation is
\be   \label{42}
\frac{\p S_{\lm,I} }{\p t}|_{\phi_l=0}=
\hf \frac{1}{\sqrt N} \int_x\int_y \dot \DD _{h}(x-y)\Big[\frac{1}{\sqrt N}J_{JI}(x)J_{JI}(y)\DD_h(x-y)+ J_{JI}(x)\dd(x-y)\dd_{IJ}\Big],
\ee
where
\[
J_{IJ}(x)=2\chi (x)\dd_{IJ} +\sqrt N 2 \p_\mu\chi^\mu_{IJ}(x)+ \sqrt N 4\chi^\mu_{IJ}(x)\frac{\p}{\p x^\mu}+\sqrt N \chi^\mu_{KI}\chi_{\mu KJ}(x).
\]

\begin{figure} 
\begin{center}
\includegraphics[width=4cm]{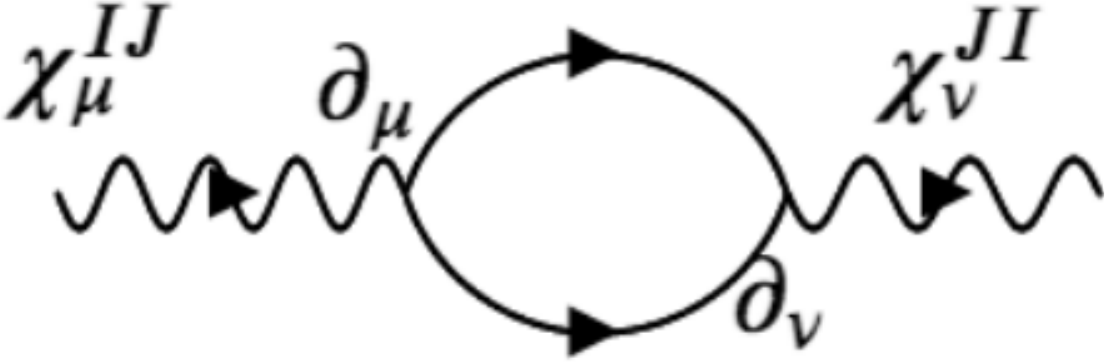}
\hspace{0.3 in}
\vspace{0.4 in}
\includegraphics[width=4cm]{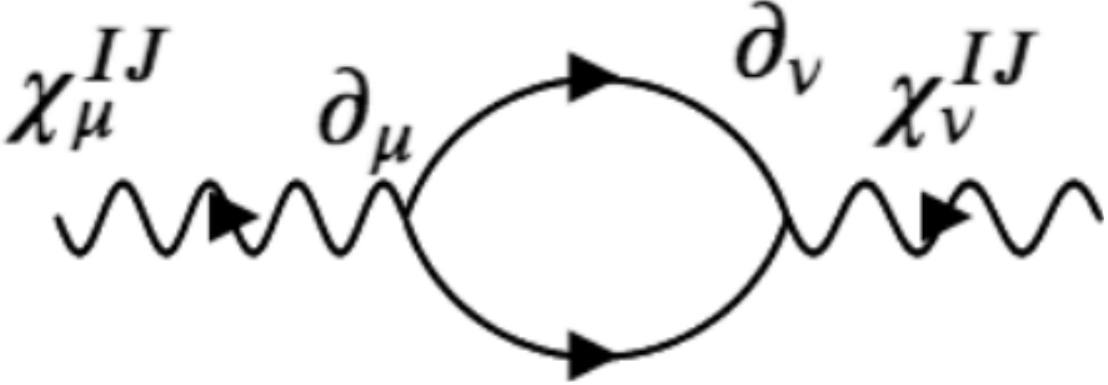}

\includegraphics[width=4cm]{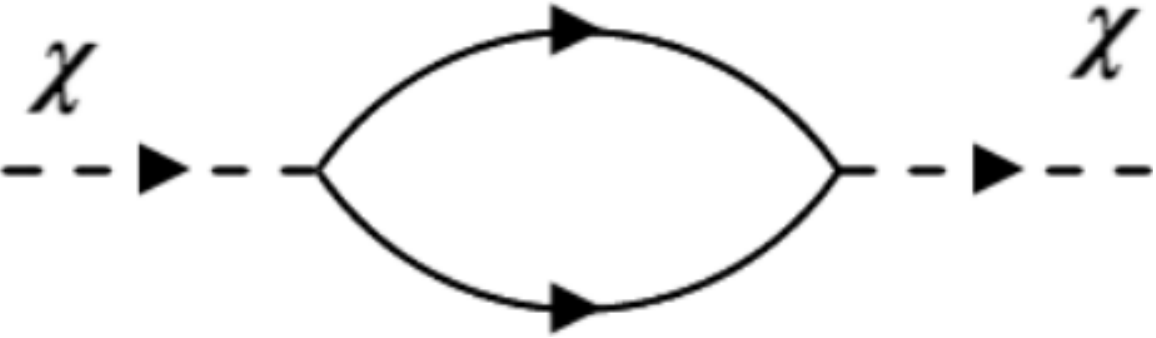}
\hspace{0.3 in}
\includegraphics[width=4cm]{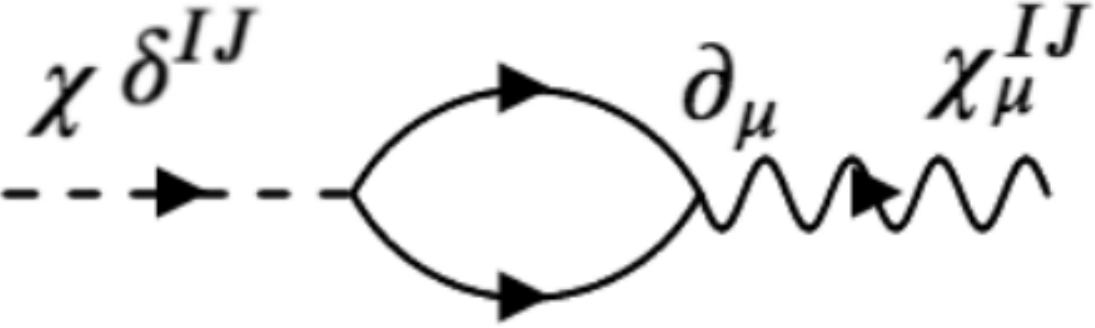}
\end{center}
\caption{Vacuum Polarization Diagrams}
\label{vacpol}
\end{figure}

These give rise to the usual one loop diagrams in Figure \ref{vacpol},
but with $\DD_h(x-y)$, the high energy propagator. 

\item {\bf Note:} The part that involves $\chi^\mu$ in \eqref{42}  is $\frac{d}{dt}$ of the following:
\[
\int_x\int_y  \DD _{h}(x-y)[J_{JI}(x)J_{JI}(y)\DD_h(x-y)]+\int_x \dd_{\mu\nu}\frac{1}{\sqrt N} \chi^\rho_{KI}(x)\chi_{\rho KI}(x)\DD_h(0).
\]
(Since $\chi^\mu_{IJ}$ is antisymmetric in $I,J,$ the $\dd_{IJ}$ term in \eqref{42} does not contribute.)

In continuum scalar QED, the two terms would combine if dimensional regularization were used and give a transverse vacuum polarization. But, with momentum cutoff schemes, the second term is a  divergent local mass term
that potentially violates gauge invariance, and can be removed by a counterterm. We do the same. Since we are working with transverse fields, we are not concerned with manifest gauge invariance. It has to be imposed by Ward Identities.  

Thus, we get for the answer the time derivative of the following vacuum polarization term:
\be  \label{46}
 \hf\int_x\int_y 16 \chi_{IJ}^\mu(x)\chi_{IJ}^\nu(y)
\Big[\frac{\p^2\DD_h(x-y)}{\p x^\mu\p y^\nu} \DD_h(x-y) -\frac{\p \DD_h(x-y)}{\p x^\mu}\frac{\p \DD_h(x-y)}{\p y^\nu}\Big],
\ee

writing in momentum space,
\be   \label{Imunu}
=-16 \hf\int_q \chi_{IJ}^\mu(q)\chi_{IJ}^\nu(-q) {\cal I}_{\mu\nu}(q,\lm),
\ee
where ${\cal I}_{\mu\nu}(q,\lm)$  
 is the standard photon vacuum polarization diagram in QED but with high energy propagators. We need to evaluate the diagrams in Figure \ref{vacpol} with a high energy propagator $\DD_h$. We are interested in the cutoff $\lm$ dependence, and we need $\lm \frac{d}{d \lm}$ of the result. We replace $\chi^\mu_{IJ}$ by $\frac{1}{i}\frac{\dd}{\dd \sigma_\mu^{IJ}}$, and this gives the leading term in the ERG equation for $S_\lm[0,\si_\mu^{TIJ}]\equiv S_\lm[\si_\mu^{TIJ}]$, where $\phi_l$ has been set to zero. 

The ERG equation is thus
\be \label{erg1}
\boxed{   
	\frac{\p}{\p t}e^{-S_\lm[\si^T_\mu]}=8\int_q\Big(  \dot {\cal I}_{\mu\nu}(q)\frac{\dd^2}{\dd \si_\mu^{TIJ}(q)\dd\si_\nu^{TIJ}(-q)}+...O(\frac{1}{\sqrt N})\Big)e^{-S_\lm[\si^T_\mu]}
}.
\ee
From now on, we use the symbol $S_\lm[\si_\mu^{TIJ}]$ for $S_{\lm,I}[0,\si_\mu^{TIJ}]$, and we have restored the superscript $T$ to indicate that we are always dealing with transverse variables.
\end{enumerate}

Let us make explicit the $\lm$ dependence by writing ${\cal I}(q^2,\lm)$. We assume that $\lo\to \infty$ for the moment. Then, ${\cal I}(q^2,0)$ is the full propagator of $\si_\mu^{IJ}$, and is fixed by dimensional analysis to be $\gamma q^{D-2}$, where $\gamma$ is a normalization factor. So, we denote by $ I(q^2)$
\be
I(q^2,\lm)= \gamma q^{D-2} - {\cal I}(q^2,\lm).
\ee
It has the property that when $\lm \to 0$, it vanishes, and when $\lm \to \infty$, it gives the full propagator. As explained in \cite{Sathiapalan:2020}, this form is more convenient to work with because it makes the calculations very similar to AdS/CFT calculations. Note that
\[
\dot {\cal I}(q^2)=-\dot I(q^2),
\]
so, the first term in the ERG equation changes sign.

We now recall that $\si_\mu^{IJ},\chi^\mu_{IJ}$ are actually
transverse, i.e., $\si_\mu^{TIJ},\chi^{T\mu}_{IJ}$, and so $ {\cal I}_{\mu\nu}(q) = \dd_{\mu\nu} {\cal I}(q^2)$.
Thus, the leading term for $S_\lm[\si_\mu^{IJ}]$ is 
\be  \label{65}
S_\lm[\si_\mu^{TIJ}] = \int_q  \hf \frac{\si_\mu ^{TIJ}(q) \si^{T\mu IJ}(-q)}{ {\cal I}(q^2)}
+O(\frac{1}{\sqrt N}).
\ee

\subsection{Evolution Operator as Functional Integral}
It is easy to apply the techniques of \cite{Sathiapalan:2017} to write the evolution operator for the ERG equation \eqref{erg1} as
\be \label{vector_evo}
e^{-S_{\lm_f}[\si^T_{\mu f}]}=\int {\cal D}\si_{\mu i}\int _{\si^T_\mu(t_i)=\si^T_{\mu i}}^{\si^T_\mu (t_f)=\si^T_{\mu f}}{\cal D}\si_\mu^{TIJ}(t)e^{-\hf\int_{t_i}^{t_f} dt\int_q \frac{1}{16 \dot I(q)} \dot \si_\mu ^{TIJ}(q)\dot \si^{T\mu IJ}(-q)}e^{-S_{\lm_i}[\si^T_{\mu i}]}.
\ee
Here, $\lm_i=e^{-t_i}, \lm_f=e^{-t_f}$. 

Hence, the Euclidean action for radial evolution is given by,
\begin{align}\label{vectoraction}
S= \hf\int_{t_i}^{t_f} dt\int_q \frac{1}{16 \dot I(q)} \dot \si_\mu ^{TIJ}(q)\dot \si^{T\mu IJ}(-q).
\end{align}

\subsection{Boundary and Bulk: Matching Degrees of Freedom}

Let us consider asymptotically AdS space with boundary at $z=0$. We would like to see how a boundary field can be continued into the AdS bulk. We take the AdS metric to be
\be
d^2= \frac{dz^2+ \dd_{\mu\nu}dx^\mu dx^\nu }{z^2}.
\ee

For the vector, there are two possibilities:
One can write an ERG equation for $S[\si_\mu]$ or for $W[A^\mu]$. Here $\si_\mu$ stands for the boundary current: 
\[
\si_\mu =\phi^I\olra\p_\mu \phi^J.
\] 
Due to current conservation, it obeys 
\[
\p^\mu \si_\mu=0.
\]

 $A^\mu$ is the source that couples to $\si_\mu$:
\[
\frac{\dd}{\dd A^\mu}Z[A^\mu]= \langle \si _\mu \rangle .
\]
This also means that $W[A^\mu]$ is gauge invariant, and so, one can choose the Lorentz gauge condition $\p_\mu A^\mu=0$. Thus, both $\si_\mu$ and $A^\mu$ can be chosen to be transverse.
\[
\si _\mu^T(p) = P_\mu ^{~\nu}(p)\si_\nu(p);
\]
\[
A^{T\mu}(p)=P^\mu_{~\nu}(p)A^\nu (p),
\]
with
\[
P_{\mu\nu}=\dd_{\mu\nu}-\frac{p_\mu p_\nu}{p^2}.
\]
Thus, in $D$ dimensions, there are $D-1$ independent components. 

Now, in the $D+1$  dimensional AdS bulk, let us assume that there is a gauge field $a^M$ with the action 
\be 
\int d^Dx \int dz z^{-D+3} (2F_{z\mu}F_{z\mu}+F_{\mu\nu}F_{\mu\nu}).
\ee
In the bulk, following \cite{Kabat}, we choose an axial gauge where $a^z=0$.  So,
\[
F_{\mu z}= \p_z a_\mu.
\]
Then the EOM are
\br
-\p^\mu (z^{3-D}\p_z a_\mu)&=&0;  \label{5.123}  \\
\p_z (z^{3-D}\p_z a_\mu + \p_\nu z^{3-D} F_{\nu\mu})
&=& 0.  \label{5.124}
\er

The equations of motion arise when we evaluate \eqref{vector_evo} in the semi-classical approximation. We have not added a source term in \eqref{vector_evo} so when mapped to AdS one expects the source free Maxwell's equation as above. \eqref{vector_evo} also implies a boundary condition:  we take $a_\mu (z=0)$ proportional to $\si_\mu$---the current of the boundary theory. Thus, $\p_\mu a_\mu(z=0)=0$.
From \eqref{5.123}, we can set $\p_\mu a_\mu=0$ everywhere in the bulk since it is zero in the boundary. This means the number of degrees of freedom is $D-1$---as it should be for a massless gauge field in $D+1$ dimensions


So, the nonzero components of $\si_\mu$ are the ones that get continued to the bulk and matches with the number of physical bulk degrees of freedom.

Then the second equation simply becomes
\be   \label{maxwell}
(\p_z^2+\Box + \frac{3-D}{z}\p_z) a_\mu=0,
\ee
where $\Box$ is the $D$ dimensional Laplacian.

Thus, to conclude, we can choose $a_\mu$ in the bulk to be equal, at the boundary, to  the current $\si_\mu$.  ERG equations can be written  for the Wilson action for $\si_\mu$.

\vspace{0.1 in}

\section{Mapping to AdS} \label{map}

We map the $D+1$ dimensional action \eqref{vectoraction} to an action in AdS space.

From \eqref{vector_evo}, we get the radial evolution action
\begin{align*}
S&=\hf\int_{t_i}^{t_f} dt\int_q \frac{1}{16 \dot I(q)} \dot \si_\mu ^{TIJ}(q)\dot \si^{T\mu IJ}(-q)\\
& = \frac{1}{2} \int_{t_i}^{t_f} \frac{dt}{e^{Dt}} \int_q \frac{\dot{\sigma}_\mu^{TIJ}(q)\dot{\sigma}^{T\mu}_{IJ}(-q)}{f^2}; ~~~ f\equiv \sqrt{16 \dot I(q)e^{-Dt}}.\\
\end{align*}

As we are always dealing with transverse variables for conveninece, we do not use the superscript T henceforth in this section.

\paragraph{Step I}
We redefine the field in terms of $f$, a smooth function of momentum and the cutoff scale:
\begin{center}
\boxed{ \sigma_\mu= \phi_\mu f}.
\end{center}
This is a field redefinition\footnote{It maps a generalized free field to a free field. See \cite{Duetsch} for a discussion of generalized fields in the context of the AdS/CFT correspondence.}.
This results in
\begin{align}
& \frac{1}{e^{tD}}\frac{\dot{\sigma}_\mu^{IJ}(q)\dot{\sigma}^{\mu}_{IJ}(-q)}{f^2}\nonumber\\
=& \frac{\dot \phi_\mu \dot \phi^\mu}{e^{tD}} + \phi_\mu \phi^\mu \left \lbrace \frac{1}{e^{tD}}\frac{\dot f^2}{f^2} -\frac{d}{dt} \left( \frac{\dot f}{f} \frac{1}{e^{Dt}} \right) \right \rbrace+ \underbrace{\frac{d}{dt} \left \lbrace \phi_\mu \phi^\mu \frac{\dot f}{f} \frac{1}{e^{Dt}} \right \rbrace}_{boundary~ term}. \label{boundaryterm}
\end{align}
 In Appendix \ref{f-c}, we show that this term does not contribute to non-analytic p-dependence in Green's functions, and can therefore be subtracted away.

\vspace{0.1 in}

So, we are left with 
\begin{align*}
S= \frac{1}{2} \int \Dp \int dt \left[ \frac{\dot \phi_\mu \dot \phi^\mu}{e^{td}}+\phi_\mu \phi^\mu \left \lbrace \frac{1}{e^{td}}\frac{\dot f^2}{f^2} -\frac{d}{dt} \left( \frac{\dot f}{f} \frac{1}{e^{dt}} \right) \right \rbrace  \right].
\end{align*}

\paragraph{Step II}
We reinterpret the scale of the boundary theory as the AdS radial coordinate.
\begin{center}
\boxed{z=e^t}.
\end{center}
\begin{align*}
S =& \frac{1}{2} \int \frac{d^D q}{z^{D+1}} \int dz  \left( z^2 \frac{\p}{\p z}\phi^\mu\frac{\p}{\p z}\phi_\mu \right) \\
&-\frac{1}{2} \int d^D q \int \frac{dz}{z} \phi^\mu \phi_\mu \left[ \frac{1}{f^2 z^{D-2} } \left( \frac{\p f}{\p z} \right)^2- z\frac{\p}{\p z}\left( \frac{\p f}{\p z}\frac{1}{f z^{D-1}} \right)\right].
\end{align*}

\paragraph{Step III}
To be able to write the kinetic term in standard AdS form, we require $f$ to satisfy the following constraint:
\begin{align}
\boxed{
\frac{\p}{\p z} \left( z^{-D+1} \frac{\p}{\p z} \frac{1}{f} \right) = z^{-D+1} \left(p^2z^2+m^2\right) \frac{1}{f}
},\label{fconstraint}
\end{align}
as in \cite{Sathiapalan:2017}, such that now
\begin{align*}
S = \frac{1}{2} \int \frac{d^D q}{z^{D+1}} \int dz  \left( z^2 \frac{\p}{\p z}\phi^\mu\frac{\p}{\p z}\phi_\mu \right)- \int \frac{d^D q}{z^{D+1}}\int dz ~~\phi_\mu \phi^\mu \left[ p^2z^2+m^2\right].
\end{align*}

$\phi_\mu$ also satisfies same equation as $\frac{1}{f}$,  i.e.,
\begin{align}\label{eqamu}
\frac{\p}{\p z} \left( z^{-D+1} \frac{\p}{\p z} \phi_\mu \right) = z^{-d+1} \left(p^2z^2+m^2\right) \phi_\mu.
\end{align}

\paragraph{Step IV} \emph{Determination of $m^2$}

\vspace{0.1 in}

The solutions of the differential equation \eqref{fconstraint} are the modified Bessel functions $I_\nu(pz)$ and $K_\nu(pz)$, with $\nu^2= m^2 +\frac{D^2}{4}$. As in \cite{Sathiapalan:2017}, we express $\frac{1}{f}$ in terms of these functions: $1/f=A(p) z^{\Dt} K_\nu (pz) + B(p) z^{\Dt} I_\nu(pz)$ for arbitrary $A(p)$ and $B(p)$.  The Green's function is also then given in terms of these functions and it behaves like $p^{-2\nu}$ at low energies. At this limit, one expects $G \rightarrow p^{2\Delta-D}$, where $\Delta$ is the conformal dimension of the boundary operator. For the vector, $\Delta= D-1$. So we get,
$\nu=\pm(1-\frac{D}{2})$, and
\begin{center}
\boxed{m^2=1-D}.
\end{center}

So, this is an action for D scalar fields whose E.O.M. are given by
\begin{align*}
z^{d-1} \frac{\p}{\p z} \left( \frac{1}{z^{d-1}} \frac{\p}{\p z} \phi^\mu \right)+ \frac{\p}{\p x_i} \frac{\p}{\p x_i} \phi^\mu- \frac{1-d}{z^2}\phi^\mu=0.
\end{align*}
This is the E.O.M which was derived in \cite{Kabat}. Going to Maxwell's equation needs one more redefinition. Note that this scalar field is equivalent to gauge field in vielbien basis as pointed out in \cite{Kabat}.

\paragraph{Step V} \emph{Obtaining Maxwell's equations}

We do a further redefinition
\begin{center}
\boxed{\phi^\mu=z a^\mu},
\end{center}
 which results in \eqref{maxwell}:
 \be   
(\p_z^2+\Box + \frac{3-D}{z}\p_z) a_\mu=0.
\ee

This concludes our derivation of Maxwell's equation as the Holographic RG equation for the conserved current perturbation.
%

One more point:
 In \eqref{boundaryterm}, we are left with a boundary term after redefining the field. For this redefinition to be unproblematic, we need to ensure two things. (i) For the redefinition itself to make sense, the function $1/f$ involved in the redefinition has to be analytic everywhere. Since the function is expressed in terms of Bessel functions, we need only check for analyticity at the boundary, as $z\to 0$. (ii) The integrand of the boundary term in the action is analytic. In fact one can show that for the values of $\nu$ discussed in this paper these constraints are satisfied.
 The analysis is given in Appendix \ref{f-c}.  

%% file: ERGtensor_input.tex
\section{ERG for Energy-Momentum Tensor}  \label{tensor}
In this section, we study a perturbation of the boundary fixed point theory by a conserved spin 2 tensor, i.e., the energy momentum tensor $T_{\mu\nu}$. We proceed as we did for the vector. We introduce an auxiliary field $\si_{\mu\nu}$ to  stand for the composite $T_{\mu\nu}$ as well as a source $h^{\mu\nu}$ for it  and write down an ERG equation for the action $S_\lm[\si_{\mu\nu}]$ for $\si_{\mu\nu}$. The evolution operator for this ERG equation is a $D+1$ dimenional field theory as before. When this is mapped to AdS space one expects to get an action for a massless spin 2 field, viz. the graviton, in the bulk. We work out only the quadratic part of the bulk action, (kinetic term), in this paper. This is intended as a proof of the principle that a dynamical graviton emerges out of Exact RG without invoking an AdS/CFT  conjecture. 

As in the case of the vector current, in the boundary CFT, we keep only the physical degrees of freedom.
The procedure for obtaining an AdS action is again the same as for a scalar field decribed in \cite{Sathiapalan:2020}.
This has the consequence that one obtains  a gauge fixed version of the quadratic graviton action in AdS space. The connection with the gauge invariant action has been worked out in the AdS/CFT literature \cite{Arutyunov,Arefeva,Liu,Mueck,Kabat} and we just follow the same steps. 

The interactions of the bulk gravitational field can in principle be worked out, but we do not attempt this in this paper. Many of the interactions should be determined by requiring manifest general coordinate or diffeomorphism invariance. Understanding these issues is work for the future.

Furthermore, for the CFT at the boundary, we consider the simplest case of a free scalar theory.
The form of the {\em quadratic} bulk graviton action is clearly independent of this choice.

We start with
\be
Z[h_{\mu\nu}]= \int \CD \phi^I e^{-\hf \int _x \p_\mu \phi^I \p^\mu \phi^I +\int_x h_{\mu\nu}\Theta^{\mu\nu}}.
\ee 
Here, $\Theta ^{\mu\nu}$ is the {\em improved} energy momentum tensor given by
\be
\Theta_{\mu\nu}= \p_\mu\phi^I \p_\nu \phi^I -\hf \dd_{\mu\nu} \p_\al \phi ^I\p^\al \phi^I - \frac{D-2}{4(D-1)} s_{\mu\nu}[\phi^2],
\ee
where
\be
s_{\mu\nu}[\phi^2]\equiv (\p_\mu\p_\nu -\dd_{\mu\nu} \Box)\phi^2
\ee
is a transverse piece that has been added to make the EM Tensor traceless.

$\Theta_{\mu\nu}$ can be rewritten as
\be
\Theta_{\mu\nu}=\frac{D}{4(D-1)}t_{\mu\nu}[\phi^2] - \phi t_{\mu\nu}\phi + \frac {D-2}{2D}\dd_{\mu\nu}\phi\Box \phi,
\ee
with $t_{\mu\nu}$ traceless:
\be
t_{\mu\nu}[\phi^2] \equiv (\p_\mu\p_\nu -\frac{\dd_{\mu\nu}}{D}\Box)\phi^2.
\ee
The last term in $\Theta_{\mu\nu}$ is not traceless but vanishes on-shell.

$\Theta_{\mu\nu}$ is conserved, (due to diffeomorphism invariance): 
\[
\p^\mu \Theta_{\mu\nu}=0.
\] 
Therefore, there is an invariance under
\[
\dd h_{\mu\nu}=\p_\mu \xi_\nu + \p_\nu \xi_\mu,
\]
which allows us to set a gauge condition
\[
\p^\mu h_{\mu\nu}=0.
\]
Thus we can define
\be
h^{T\nu}_\mu=P_\mu^{~\rho}P_\si^{~\nu}h_\rho^{~\si},
\ee
with
\[
P_{\mu\nu}=\dd_{\mu\nu}-\frac{\p_\mu \p_\nu}{\Box}.
\]

It is also traceless because it is a CFT:
\[
\Theta^\mu_\mu=0.
\]
So, we can choose $h^\mu_\mu=0$.

\subsection{Auxiliary Field}

Now we can define an auxiliary field $\si^{\mu\nu}$ by
\be
\si^{\mu\nu}= \Theta^{\mu\nu}
\ee
in the bare theory. Thus we can introduce it in the functional integral by
\[
\int \CD \si^{\mu\nu} \dd (\si^{\mu\nu}-\frac{\dd}{\dd h_{\mu\nu}})Z[h_{\mu\nu}]= Z[h_{\mu\nu}]
\]
\[=
\int \CD \chi _{\mu\nu} \int \CD \si^{\mu\nu} e^{i\int _x \chi_{\mu\nu}(\si^{\mu\nu}-\frac{\dd}{\dd h_{\mu\nu}})}Z[h_{\mu\nu}],
\]
(note that $\chi_{\mu\nu}$ can also be taken to be transverse and traceless since its trace and longitudinal components do not couple to anything, since $\si_{\mu\nu}$ is traceless and transverse),
\begin{align*}
	=&
	\int \CD \si^{\mu\nu}\int \CD \chi _{\mu\nu}  e^{i\int _x \chi_{\mu\nu}\si^{\mu\nu} }Z[h_{\mu\nu}-i\chi_{\mu\nu}]\\
	=&
	\int \CD \si^{\mu\nu}e^{\int_x h_{\mu\nu}\si^{\mu\nu}}\int \CD \chi _{\mu\nu}  e^{i\int _x \chi_{\mu\nu}\si^{\mu\nu} }Z[-i\chi_{\mu\nu}].
\end{align*}
Thus, we can write
\be
Z[h_{\mu\nu}]=e^{W[h_{\mu\nu}]}=\int \CD \si^{\mu\nu}e^{-S[\si^{\mu\nu}] +\int_x h_{\mu\nu}\si^{\mu\nu}}.
\ee
This defines
$S[\si_{\mu\nu}]$, which is a useful object---it encodes information about the correlators of $\Theta^{\mu\nu}$:
\be
\int \CD \si_{\mu\nu} ~\si_{\mu_1 \nu_1} (x_1)...\si_{\mu_n,\nu_n}(x_n) e^{-S[\si_{\mu\nu}]}=
\langle \Theta_{\mu_1 \nu_1} (x_1)...\Theta_{\mu_n,\nu_n}(x_n)\rangle.
\ee
$W[h_{\mu\nu}]$ is the usual generator of connected correlations. One can write ERG equations for both $W$ and $S$. Let us proceed to write an ERG equation for $S[\si_{\mu\nu}]$.

\subsection{ERG equation for $S[\si_{\mu\nu}]$}

We start by writing
\begin{align}
	e^{-S[\si_{\mu\nu}]}&=\int \CD \chi _{\mu\nu}  e^{i\int _x \chi_{\mu\nu}\si^{\mu\nu} }Z[-i\chi_{\mu\nu}]\nonumber\\
	&\label{S1}
	=
	\int \CD \chi _{\mu\nu}  e^{i\int _x \chi_{\mu\nu}\si^{\mu\nu} }\int \CD \phi^I e^{-\hf \int _p  \phi^I(p) \DD^{-1}(p^2) \phi^I(-p) -i\int_p \chi_{\mu\nu}(p)\Theta^{\mu\nu}[\phi^I](-p)}.
\end{align}
Now write $\phi^I= \phi^I_l +\phi^I_h$ and
$\DD(p^2)=\frac{K_0}{p^2}= \DD_l+\DD_h$, or,
\[
\frac{K_0}{p^2}=  \frac{K}{p^2}+ \frac{K_0-K}{p^2}.
\]
As in previous sections, the functions $K_0=e^{-\frac{p^2}{\lo^2}}$ and $K=e^{-\frac{p^2}{\lm^2}}$ represent a typical choice but as always $\DD,\DD_l,\DD_h$ can be very general functions.
$\lo$ is UV cutoff that can be taken to be $\infty$, and $\lm$ is the moving IR cutoff.
\begin{align}
	e^{-S[\si_{\mu\nu}]}=&
	\int \CD \phi_l^I e^{-\hf \int _p  \phi_l^I(p) \DD_l^{-1}(p^2) \phi_l^I(-p)}\int \CD \chi _{\mu\nu}  e^{i\int _x \chi_{\mu\nu}\si^{\mu\nu} } \times\nonumber\\
	&\int \CD \phi_h^I e^{-\hf \int _p  \phi_h^I(p) \DD_h^{-1}(p^2) \phi_h^I(-p)-i\int_p \chi_{\mu\nu}(p)\Theta^{\mu\nu}[\phi_l^I+\phi_h^I](-p)},\\
	e^{-S[\si_{\mu\nu}]}=&
	\int \CD \phi_l^I e^{-\hf \int _p  \phi_l^I(p) \DD_l^{-1}(p^2) \phi_l^I(-p)}e^{-S_\lm[\si_{\mu\nu},\phi_l^I]},
\end{align}
with
\be 
e^{-S_\lm[\si_{\mu\nu},\phi_l^I]}=\int \CD \chi _{\mu\nu}  e^{i\int _x \chi_{\mu\nu}\si^{\mu\nu} }\int \CD \phi_h^I e^{-\hf \int _p  \phi_h^I(p) \DD_h^{-1}(p^2) \phi_h^I(-p)-i\int_p \chi_{\mu\nu}(p)\Theta^{\mu\nu}[\phi_l^I+\phi_h^I](-p)}.
\ee

We obtain the following ERG equation of $S[\si_{\mu\nu}]$ to the lowest order in $1/N$ like in the vector case, (details are in Appendix \ref{tensorAppendix}): 
\begin{center}
\boxed{
\frac{\p}{\p t}e^{-S_\lm[\si_{\mu\nu}]}=
-\hf \int_x\int_y 4 \dot \DD_h(x-y)\p_x^\mu\p_x^\nu \p_y^\rho\p_y^\si\DD_h(x-y)\frac{\dd^2}{\dd \si^{\mu\nu}(x)\dd \si^{\rho\si}(y)}e^{-S_\lm[\si^{\mu\nu}]}
}.
\end{center}

Writing in momentum basis,
\begin{equation}
\frac{\p}{\p t}e^{-S_\lm[\si_{\mu\nu}]}=
-\hf \int_q \dot{\mathcal{I}}_{\mu\nu\rho\sigma}(q,\Lambda)\frac{\dd^2}{\dd \si_{\mu\nu}(q)\dd \si_{\rho\si}(-q)}e^{-S_\lm[\si^{\mu\nu}]}.\label{momergtensor}
\end{equation}

It is easy to see using transversality and tracelessness of $\chi_{\mu\nu}$ (and $\si^{\mu\nu}$) that the only tensor structure that contributes to ${\cal I}_{\mu\nu\rho\si}$ is of the form
$\dd^{\mu\rho}\dd^{\nu\sigma}+\dd^{\nu\rho}\dd^{\mu\sigma}$. See Appendix \ref{thetamunu} for the form of the two point function ${\cal I}_{\mu\nu\rho\si}(q)$.

As discussed in the vector case, to make the propagator vanish at $\Lambda \rightarrow 0$, we define new propagator as
\be
I(q^2,\lm)= \gamma q^{D} - {\cal I}(q^2,\lm),
\ee
\[
\dot {\cal I}(q^2)=-\dot I(q^2).
\]
Like before, the first term in the ERG equation changes sign.

\color{black}

Thus, we get in momentum space
\be
\frac{\p}{\p t}e^{-S_\lm[\si^{\mu\nu}]}=\hf\int_p
\dot I(p^2) \frac{\dd^2}{\dd \si^{\mu\nu}(p)\dd \si_{\mu\nu}(-p)}e^{-S_\lm[\si^{\mu\nu}]}~+~ higher~order.
\ee
The precise expression for $I(p^2)$ depends on the regulator. Note that raising and lowering indices is done with the flat space Euclidean metric. If we denote the composite scalar propagator in momentum space by
\be
G_{scalar}(p^2)= \int \Dp e^{ip.X}\DD_h(X)^2
\ee 
then
\be  \label{ts}
G_{tensor}(p^2)=I(p^2)\approx (p^2)^2 G_{scalar}(p^2).
\ee

The evolution operator for this ERG equation is known, and we get
\be
e^{-S_\lm[\si^{\mu\nu}_f]}=\int \CD \si^{\mu\nu}_i(p)\int _{\si^{\mu\nu}(p,t_i)=\si^{\mu\nu}_i(p)}^{\si^{\mu\nu}(p,t_f)=\si^{\mu\nu}_f(p)}\CD \si^{\mu\nu}(p,t)
e^{-\hf\int_p \int_{t_i}^{t_f}dt~\frac{\dot \si^{\mu\nu}(p,t)\dot \si_{\mu\nu}(p,t)}{\dot I(p^2)}}e^{-S_{\lo}[\si^{\mu\nu}_i]}.
\ee
Here, we take $e^{-t_i}=\lo,~e^{-t_f}=\lm$ in some units.

The action of radial evolution is then 
\begin{align}\label{tensoraction}
S= \hf \int_p \int_{t_i}^{t_f}dt~\frac{\dot \si^{\mu\nu}(p,t)\dot \si_{\mu\nu}(p,t)}{\dot I(p^2)}.
\end{align}

\subsection{Matching Bulk and Boundary Degrees of Freedom}
Let us now consider a conserved energy-momentum tensor $T_{\mu\nu}$ in a CFT satisfying
\br
\p^\mu T_{\mu\nu}&=&0;\\
T^\mu_{~\mu}&=&0.
\er
It can be defined by introducing a background metric $g_{\mu\nu}=\dd_{\mu\nu}+h_{\mu\nu}$. Then,
\[
\frac{\dd}{\dd h_{\mu\nu}}Z[h_{\mu\nu}]=\langle T^{\mu\nu}\rangle.
\]
We consider an auxiliary field, as before,
defined by
\[
\si^{\mu\nu}=\frac{\dd}{\dd h_{\mu\nu}}
\]
so that the coupling is $h_{\mu\nu}\si^{\mu\nu}$. Since
\[
\p^\mu T_{\mu\nu}=0,
\]
we can use diffeomorphism invariance to choose 
\be   \label{5.2.5}
\p^\mu h_{\mu\nu}=0.
\ee
$h_{\mu\nu}$ has $\hf D(D+1)$ d.o.f. The transversality condition makes it $\hf D(D-1)$. It being a CFT, we can also set the trace to zero:
\be  \label{5.2.6}
h^\mu_{~\mu}=0
\ee
Thus, in $D=3$, we get a traceless symmetric $2\times 2$ matrix which has two d.o.f. If this continues to the four dimensional bulk, it should become two physical propagating degrees---which is correct for a massless spin 2 graviton in four dimensions.

\vspace{0.1 in}
To check this, let us consider the linearized equations for $h_{MN}$ in the gauge $h_{zM}=0$.
\br
\Box h - \p_\mu\p^\nu h^\mu_\nu+\frac{1-D}{z}\p_z h&=&0;\label{5.2.7}\\
\p_z(\p_\mu h-\p_\nu h^{~\nu}_\mu)&=&0\label{5.2.8};\\
\p_z^2h^{~\mu}_\nu+\Box h^{~\mu}_\nu +\frac{1-D}{z}\p_z h^{~\mu}_\nu -\frac{1}{z} \p_z h \dd^\mu_\nu -\p^\mu\p_\rho h^\rho_{~\nu}-\p_\nu\p^\rho h_{~\rho}^{~\mu}&=&0. \label{5.2.9}
\er
The first two equations are constraints. 

We would like to show that the conditions \eqref{5.2.5} and \eqref{5.2.6} continue to hold in the bulk. To this end, we follow \cite{Arutyunov}.

\vspace{0.1 in}
Define a transverse tensor:
\be
h^{T\nu}_\mu=P_\mu^{~\rho}P_\si^{~\nu}h_\rho^{~\si},
\ee
\[
P_\mu^{~\rho}= \dd_{\mu}^{~\rho} - 
\frac{\p_\mu \p^\rho}{\Box}.
\]
Then, its trace is
\be \label{5.2.11}
h^T=h- \frac{\p^\nu\p_\mu h_\nu^{~\mu}}{\Box}.
\ee
So,
\be  \label{5.2.12}
 (h-h^T)= \frac{\p^\nu\p_\mu h_\nu^{~\mu}}{\Box}.
\ee
Acting with $\p^\mu$ on \eqref{5.2.8} implies that
\be
\p_z( h-\frac{\p^\nu\p_\mu h_\nu^{~\mu}}{\Box}){\Box}=0 \implies \p_z  h^T=0. \label{5.2.13}
\ee

On the boundary, $h=h^T$. Also, we have chosen $h=0$ on the boundary. Thus, $h^T=0$ on the boundary. Then, \eqref{5.2.13} implies that $h^T=0$ everywhere.

Now \eqref{5.2.7} implies that, (using \eqref{5.2.11}), 
\be
\Box h^T=0= \frac{D-1}{z} \p_z h.
\ee
This means $h=0$ everywhere given that it is zero on the boundary.
Then, \eqref{5.2.8} implies that 
\be
\p_z \p_\nu h_\mu^{~\nu}=0,
\ee
and since $\p_\nu h_\mu^{~\nu}=0$ on the boundary, it is zero everywhere. Thus, we have shown that the conditions 
$h= \p_\nu h_\mu^{~\nu}=0$ holds everwhere in the bulk (in the gauge $h_{zM}=0$),
i.e.,
\be 
h_{\mu\nu}=h^T_{\mu\nu}~~;~~h=0~~~;~~~h_{zM}=0
\ee
everywhere. This is what was to be shown. 

If we use these conditions, \eqref{5.2.9} becomes
\be
\p_z^2h^{~\mu}_\nu+\Box h^{~\mu}_\nu +\frac{1-D}{z}\p_z h^{~\mu}_\nu  =0.
\ee
This can be obtained from a $D+1$ dimensional AdS action
\be   \label{hmunuaction}
\int d^{d}xdz~ z^{-D+1}[\p_z h^{T\mu}_\nu \p_z h_\mu^{T\nu} +
\p_\rho h^{T\mu}_\nu \p^\rho h_\mu^{T\nu}].
\ee
We have restored the superscript $T$ to indicate that these are the transverse degrees of freedom. It is also traceless.
This is precisely the action we obtain in the ERG evolution operator in the next section.

\subsection{Mapping to AdS}

The radial evolution action \eqref{tensoraction} is that of $\frac{D(D-1)}{2}-1$ scalars and we can use the same map as in \cite{Sathiapalan:2017,Sathiapalan:2020} and in the previous section, to map this to AdS: 
\be    \label{ftensor}
\si_{\mu\nu}(p,t)= y_{\mu\nu}(p,t)f(p,t),
\ee
with $f^2= -\dot I z^{-D}$ and 
\be   \label{ftensor1}
\frac{1}{f}= z^\Dt (\al K_\nu(pz)+\beta I_\nu(pz)).
\ee
The Green's function takes the form parametrized by $\nu$ and a parameter $\gamma$:
\be   \label{Gnu}
G_\nu (pz)= \frac{\gamma p^{-\nu} K_\nu(pz)}{p^\nu (K_\nu(pz) - \frac{1}{\gamma}I_\nu(pz))}.
\ee
The Green's function in the Wilson action then behaves
as $p^{-2\nu}$ at low energies (compared to the cutoff).
Thus, for $\si^{\mu\nu}$, one expects $p^D$ since it is the energy momentum tensor. Thus, $\nu=-\Dt$. Then, since $m^2+(\Dt)^2=\nu^2$, we see that $m^2=0$. We need to choose
\be
I(p^2)=G_{tensor}(p^2)=G_{-\Dt}(p^2).
\ee
This maps the system to an action
\be   \label{sads}
S_{AdS}=\int dz \Dp z^{-1+D}\Big[\frac{\p y_{\mu\nu}(p)}{\p z}\frac{\p y_{\rho\sigma}(-p)}{\p z}+ p^2y_{\mu\nu}(p)y_{\rho\si}(-p)\Big]\dd^{\mu\rho}\dd^{\nu \si},
\ee
which should be compared to \eqref{hmunuaction}.

By defining $h_{\mu\nu}^B z^2=y_{\mu\nu}$, one obtains an action for $h^B_{\mu\nu}$, the bulk metric perturbation:
\begin{align}
	\int dz\Dp ~z^{1-D}\Big[&z^4 \bigg(\frac{\p h^B_{\mu\nu}(p)}{\p z}\frac{\p h^B_{\rho\sigma}(-p)}{\p z}+ p^2h^B_{\mu\nu}(p)h^B_{\rho\si}(-p)\bigg)\nonumber\\
	 &+ 4 z^3h^B_{\mu\nu}\frac{\p h_{\rho\si}^B}{\p z}+ 4 z^2 h_{\mu\nu}^Bh_{\rho\si}^B\Big]\dd^{\mu\rho}\dd^{\nu \si},
\end{align}
 and an equation of motion as also found in \cite{Kabat}:
 \be
(\p_z^2 -p^2)h_{\mu\nu}^B(p) +\frac{5-D}{z} \p_z h_{\mu\nu}^B(p) - \frac{2D-4}{z^2} h_{\mu\nu}^B(p)=0.
\ee
One can also define
\[ h^{B\mu}_{~~~\nu}= z^2\dd^{\mu\rho}h^B_{\rho\nu} =\delta^{\mu\rho} y_{\rho\nu}
\]
to obtain a physically equivalent equation of motion:
\be
\p_z^2h^{~\mu}_\nu -p^2 h^{~\mu}_\nu +\frac{1-d}{z}\p_z h^{~\mu}_\nu  =0,
\ee
the same as the ``scalar" -$y_{\mu\nu}$-  equation of motion obtained from 
the action \eqref{sads}. This form of the equation for the bulk metric perturbation was obtained in \cite{Arutyunov}, and also in the last subsection.

\subsection{Simultaneous Mapping of Scalar, Vector and Tensor Composites}\label{simulmap}

In the previous section, the free gauge fixed equation of motion for a (massless) graviton in AdS space was obtained by the mapping \eqref{ftensor}, where the function $f$, (let us refer to it as $f_{tensor}$), is chosen so that $f_{tensor}^2 =-\dot I z^{-D}$, and also satisfies \eqref{ftensor1}. Thus, the Green function $I$
 (referred to in the appendix 
  as $G_{tensor}$) is constrained. But the Green function is also related to the elementary scalar $\phi$-propagator:
\be
G_{tensor} (x-y) = \Delta_h(x-y)(\p_x^2)^2 \Delta_h(x-y).
\ee

Thus, based on this\footnote{The propagator at low energies is $\approx p^{-2\nu}$}, we chose $\nu =-\Dt$.
This results in the constraint:
\be  \label{tconstraint}
G_{tensor}(x-y)\equiv\Delta_h(x-y)(\p_x^2)^2 \Delta_h(x-y)=G_{-\Dt}(x-y),
\ee
where $G_\nu$ was defined in momentum space in \eqref{Gnu}.


On the other hand, by similar arguments, $f_{scalar}$ and $f_{vector}$ are also related to the $\phi$-propagator  $\Delta_h(p)$.

In each case,
 \be   \label{f}
\frac{1}{f}= z^\Dt (\al K_\nu(pz)+\beta I_\nu(pz))
\ee
with $\nu = -\Dt, -\Dt+1, -\Dt+2 $. 
Also, we need to impose:
\be  \label{vconstraint}
G_{vector}(x-y) \equiv \Delta_h(x-y)\p_x^2 \Delta_h(x-y)=G_{-\Dt+1}(x-y),
\ee
and
\be   \label{sconstraint}
G_{scalar}(x-y) \equiv \Delta_h(x-y)\Delta_h(x-y)=G_{-\Dt+2}(x-y).
\ee

The constraints \eqref{tconstraint},\eqref{vconstraint} and \eqref{sconstraint} are clearly not mutually compatible: $G_{tensor}(p),G_{vector}(p), G_{scalar}(p)$ differ only in having extra powers of $p$, as given in \eqref{ts}, whereas $G_\nu(p)$ with different $\nu$ are different Bessel functions. Thus, as things stand, only their low energy behaviours are consistent with the constraints.

The way out of this contradiction is to perform a field redefinition of $\sigma_\mu, \sigma_{\mu\nu}$ so that their kinetic terms are consistent. 
We retain \eqref{sconstraint} as the only constraint, and use it to fix $\DD(x-y)$, the propagator of the elementary scalar.

Thus, consider the functions
\br
f_v(p/\lm) &=& \bigg(\frac{\dot G_{vector}(p)}{\dot G_{-\Dt+1}(p)} \bigg)^\hf,\\
f_t(p/\lm) &=& \bigg(\frac{\dot G_{tensor}(p)}{\dot G_{-\Dt}(p)}\bigg)^\hf.
\er
Both functions have the small $p$ expansions of the form
\[
f_{v,t} (p/\lm)\approx 1+ O((p/\lm)^2).
\]
Thus, near the boundary, where $\lm\to \lo \to\infty$ for the bare continuum theory, they are equal to 1. Let us define: 
\br
\si_\mu(p) &=& f_v(p/\lm) \si'_\mu(p), \label{1.4.41} \\ \label{1.4.42}
\si_{\mu\nu}(p)&=& f_t(p/\lm)\si'_{\mu\nu}(p).
\er
This is equivalent to modifying the auxiliary fields in the bare theory by terms of the order of $\frac{p^2}{\lo^2}$. So, in the continuum theory, it reduces to the respective currents. From the point of view of the ERG, this is equivalent to a modification to a scheme for coarse graining which is controlled by the cutoff dependence of the Green function. In the limit $\lm \to 0$, the physics is scheme independent.
Then, the leading terms in the ERG equations become
\be
\frac{\p}{\p t}e^{-S_\lm[\si'^{\mu\nu}]}=\hf\int_p
\dot G_{-\Dt}(p) \frac{\dd^2}{\dd \si'^{\mu\nu}(p)\dd \si'_{\mu\nu}(-p)}e^{-S_\lm[\si'^{\mu\nu}]}~ +~ higher~order,
\ee
and
\be
\frac{\p}{\p t}e^{-S_\lm[\si'^{\mu}]}=\hf\int_p
\dot G_{-\Dt+1}(p) \frac{\dd^2}{\dd \si'^{\mu}(p)\dd \si'_{\mu}(-p)}e^{-S_\lm[\si'^{\mu}]}~ +~ higher~order.
\ee
The kinetic terms in the evolution operator become:
\be
S_{tensor}=\hf\int_p \int_{t_i}^{t_f}dt~\frac{\dot \si^{\mu\nu}(p,t)\dot \si_{\mu\nu}(p,t)}{\dot G_{-\Dt}(p^2)}
\ee
for the tensor, and
\be
S_{vector}=\hf\int_p \int_{t_i}^{t_f}dt~\frac{\dot \si^{\mu}(p,t)\dot \si_{\mu}(p,t)}{\dot G_{-\Dt+1}(p^2)}
\ee
for the vector. (We have dropped the primes for simplicity.)
Now, all kinetic terms are of the correct form and the mapping to AdS can be done consistently for all three equations.  It goes without saying that this additional field redefinition steps such as \eqref{1.4.41}, \eqref{1.4.42} need to be done for all higher spins also.

Thus, in this section we have obtained from the ERG analysis, in a manner consistent with the scalar composite mapping in \cite{Sathiapalan:2020}, and vector composite mapping given in the earlier sections, an action for the graviton  which is a gauge fixed version of the quadratic part of  Einstein's action for metric perturbations about an AdS background. Thus, one can say that dynamical bulk gravity in AdS space emerges from an ERG equation, (in the large N limit), of a boundary CFT using a well defined mathematical map. Correctness of the result is built into it because it follows from (correctness of) the ERG equation---it does not require an invocation of the AdS/CFT conjecture.  This method can be applied to interactions as well, and correctness of the boundary correlations derived from the bulk action is once again guaranteed by the same logic. The details of interactions and their locality properties have to be worked out and this is work for the future. A conceptual issue that needs to be understood is diffeomorphism invariance of the bulk from the point of view of the ERG in the boundary---this is also work for the future.


%% file: Summary_input.tex
\section{Summary and Conclusions}\label{con}

In the AdS/CFT correspondence, a massless gauge field in the bulk is dual to a conserved vector current in the boundary theory. Simlarly, the metric perturbation is dual to the energy momentum tensor in the boundary theory.
Thus, it is expected that Maxwell equations should describe the RG evolution of the boundary theory when perturbed by a currrent operator, and Einstein equations for metric perturbation describe the RG evolution of the boundary theory when perturbed by the energy momentum tensor.

In this paper, we have derived this using the techniques developed in \cite{Sathiapalan:2017}---at the linearized level---starting from the Exact RG equation of the $D$ dimensional boundary theory, which for us is the O(N) scalar model in three dimensions. This is a three step process. The first step is to write the Polchinksi ERG equation for the Wilson action of the relevant field. The second step is to write down the evolution operator for this equation in the form of a functional integral of a $D+1$ dimensional field theory---the extra dimension being the RG scale. The third step is to perform a field redefinition to map this action to AdS space.

On performing these steps one finds
that if $a_\mu(z,x)$ is the bulk gauge field, and $J_\mu(x)$ is the boundary current, then, near the boundary $z=0$,
\[ a_\mu(z,x) \approx z^{D-2} J_\mu(x),
\]
and for the bulk metric perturbation $h_{\mu\nu}(x)$ and boundary energy momentum tensor $T_{\mu\nu}$,
\[
h_{\mu\nu}(z,x)\approx z^{D-2} T_{\mu\nu}(x).
\]
Furthermore, the AdS bulk field theory functional integral is the Exact RG evolution operator of the boundary (Wilson) action for the auxiliary fields that stand for the current and energy momentum tensor respectively.

 In order to perform these steps, it was found convenient to introduce auxiliary fields for the composite operators, following the idea introduced in \cite{Sathiapalan:2020} for scalar composites. Thus, a spin one field $\si_\mu$ was introduced for the vector current, and a spin two field $\si_{\mu\nu}$ was introduced for the energy momentum tensor. A brief introduction to computations using  auxiliary fields, (standing for scalar composite operators), and ERG equations for their Wilson actions, has also been given in Section 2 and Section 3.
 These techniques should be useful in dealing with composite operators in field theory.

 Following these same techniques, Wilson actions were defined for these spin one and spin two auxiliary fields, and ERG equations derived. The ERG equation for auxiliary (composite) field actions has a Polchinski equation form but with additional terms. These additional terms map to interaction terms in the bulk theory. We discussed this for the scalar composite in the free theory. In the case of the conserved currents, the leading term maps, after performing the three steps mentioned above, to a kinetic term---describing linearized metric perturbations for spin $2$ and Maxwell equations for spin $1$. The entire computation was done after making a gauge choice. Doing all this while maintaining manifest gauge invariance is an open problem that needs to be addressed.

 It is important to point out that these steps follow logically from the ERG of the boundary theory, and do not require the AdS/CFT conjecture for their justification at any point. Thus, one can say that at least at the {\em linearized} level, one can {\em derive} actions for gauge fields and gravity in a bulk AdS theory  starting from ERG equations of a  boundary CFT. This provides some additional insight into the workings of holography.
 
 Interactions terms in the bulk action for scalar fields were discussed in \cite{Sathiapalan:2020}. This needs to be worked out for gauge fields and metric perturbations.  It should be pointed out that because all these follow from the ERG equations, they are guaranteed to give the right boundary correlations. This should be true to any order, without invoking the AdS/CFT conjecture. The real issue that needs to be understood then are the locality properties of the bulk interactions obtained this way. This is work for the future.
 
\vspace{1cm} 
 {\bf Acknowledgements:} B.S. would like to thank Hidenori Sonoda for useful discussions.

%% file: AnomDimAppendix_input.tex
\section{Anomalous Dimension of $\phi^2$ Using Auxiliary Fields}\label{anom}

In this appendix, some of the details of the anomalous dimension calculation in Section \ref{scalar} are given.

\subsection{Leading order} Each loop in eq. \eqref{lead_cor} gives a factor:
\[
\int \Dp \frac{2}{(p^2+m^2)((p+k)^2+m^2)} = \int _0^1dx~\int \Dp \frac{2}{(p^2+M^2)^2} = \frac{2}{(4\pi)^{2-\eps}}\int _0^1dx~\frac{\Gamma(\eps)}{(M^2)^\eps},
\]
where $M^2= k^2x(1-x)+m^2$, and $D=4-2\eps$. 
Using $\Gamma(\eps) = \frac{1}{\eps}+O(1)$, we get
\[
\frac{2}{(4\pi)^{2-\eps}}\frac{\Gamma(\eps)}{(M^2)^\eps} =\frac{2}{(4\pi)^2}\Big[\frac{1}{\eps} - \ln \frac{k^2x(1-x)+m^2}{4\pi \mu^2}\Big]=\frac{1}{(4\pi)^2} \Big[\frac{2}{\eps}+f(k)\Big],
\]
where $f(k)$ is the finite part. Thus, considering the one loop and two loop terms, we get
\[
\frac{1}{(4\pi)^2}\Big(\frac{2}{\eps}+f(k)\Big)+ \bar \la \Big[\frac{1}{(4\pi)^2}\Big(\frac{2}{\eps}+f(k)\Big)\Big]^2.
\]
We can add counter-terms to remove the momentum independent terms. That gives for the wave function renormalization 
\[
\langle\sigma \sigma\rangle=\frac{f}{(4\pi)^2} \Big[1+ \frac{4}{(4\pi)^2}\frac{\bar \la}{\eps}\Big]=\frac{f}{(4\pi)^2} Z.
\]
We have kept the log divergence only. 
Noting that $\frac{1}{\eps}$ pole corresponds to a divergence of the form $\ln \lm^2 \approx 2t$, we see that
\[
\sqrt Z = 1+ \frac{4\bar \la}{(4\pi)^2} t.
\]
From which, we get \eqref{lead_anom}.

\subsection{Sub-leading or $\pmb{O(1/N)}$ correction}\label{app:sublead}
Here, we detail how to obtain \eqref{1/N}.

We have, from \eqref{Zineta},
\be
Z[J]= \int {\cal D} \eta' e^{-\hf N Tr \ln \big[1+2 \frac{\DD J}{\sqrt N} - 2 \frac{\sqrt {\bar \la} \DD \eta'}{\sqrt N}\big]-\int \frac{\eta'^2}{2}},
\ee
expanding the log in powers of $\sqrt {\bar\la}$,
\begin{align*}
	=&
	\int {\cal D}\eta'\exp\bigg\{-\frac{N}{2} Tr \ln \Big[1+\frac{2\DD J}{\sqrt N}\Big] + \hf N Tr \bigg[\bigg(\frac{1}{1+\frac{2\DD J}{\sqrt N}}\bigg)2 \frac{\sqrt {\bar \la}}{\sqrt N}\DD \eta'\bigg]\\ &+\frac{N}{4} Tr \bigg[\bigg(\frac{1}{1+\frac{2\DD J}{\sqrt N}}\bigg)\frac{2\sqrt {\bar \la}}{\sqrt N}\DD \eta'\bigg(\frac{1}{1+\frac{2\DD J}{\sqrt N}}\bigg)\frac{2\sqrt {\bar \la}}{\sqrt N}\DD \eta'\bigg]\bigg\}e^{-\int \frac{\eta'^2}{2}}.
\end{align*}
We expand the exponential to $O({\bar \la})$:
\begin{align*}
	&\int {\cal D}\eta'e^{-\frac{N}{2} Tr \ln \big[1+\frac{2\DD J}{\sqrt N}\big]}\bigg\{1+\underbrace{\hf N Tr \Big[\Big(\frac{1}{1+\frac{2\DD J}{\sqrt N}}\Big)2 \frac{\sqrt {\bar \la}}{\sqrt N}\DD \eta'\Big]}_{(a)}\\
	&+\underbrace{\frac{1}{2!}\Big(\frac{N}{2}\Big)^2\bigg(Tr \bigg[\bigg(\frac{1}{1+\frac{2\DD J}{\sqrt N}}\bigg)2 \frac{\sqrt {\bar \la}}{\sqrt N}\DD \eta'\bigg]\bigg)^2}_{(b)}\\
	&+\underbrace{\frac{N}{4} Tr \bigg[\bigg(\frac{1}{1+\frac{2\DD J}{\sqrt N}}\bigg)\frac{2\sqrt {\bar \la}}{\sqrt N}\DD \eta'\bigg(\frac{1}{1+\frac{2\DD J}{\sqrt N}}\bigg)\frac{2\sqrt {\bar \la}}{\sqrt N}\DD \eta'\bigg]}_{(c)}\bigg\}e^{-\int \frac{\eta'^2}{2}}.
\end{align*}
$(a)$ gives zero because of the odd integrand $\eta'$.
We expand $(b),(c)$ to quadratic order in $J$. Each one gives two kinds of terms. They all correspond to two loop Feynman diagrams. One in each of (b),(c) is a convergent loop multiplied by a divergent self energy correction. This does not contribute to anomalous dimension of $\sigma$. 

 Ignoring the disconnected diagrams we collect $J^2$ dependent part at order $\bar \la$ :
\begin{align*}
	&\int {\cal D}\eta^\prime \bigg\lbrace \frac{1}{2!}\bigg(\frac{N}{2}\bigg)^2\bigg(Tr \bigg[\bigg(\frac{1}{1+\frac{2\DD J}{\sqrt N}}\bigg)2 \frac{\sqrt {\bar \la}}{\sqrt N}\DD \eta'\bigg]\bigg)^2\\
	&+\frac{N}{4} Tr \bigg[\bigg(\frac{1}{1+\frac{2\DD J}{\sqrt N}}\bigg)\frac{2\sqrt {\bar \la}}{\sqrt N}\DD \eta'\bigg(\frac{1}{1+\frac{2\DD J}{\sqrt N}}\bigg)\frac{2\sqrt {\bar \la}}{\sqrt N}\DD \eta'\bigg]\bigg\rbrace e^{-\int \frac{\eta'^2}{2}}\\
	=&4 \bar \la \int {\cal D}\eta^\prime  Tr(\Delta \eta^\prime) Tr [ (\Delta J)^2 \Delta \eta^\prime]e^{-\int \frac{\eta'^2}{2}}+2{\bar \la} \int {\cal D} \eta' \{ Tr (\DD J \DD \eta') Tr (\DD J \DD \eta') e^{-\int \frac{\eta'^2}{2}}\\
	&+ \frac{8\bar \la}{N} \int {\cal D}\eta^\prime Tr [ \Delta \eta^\prime (\Delta J)^2 \Delta \eta^\prime ]e^{-\int \frac{\eta'^2}{2}}+ \frac{4\bar \la}{N} \int {\cal D}\eta^\prime Tr [\Delta J  \Delta \eta^\prime (\Delta J \Delta \eta^\prime ]e^{-\int \frac{\eta'^2}{2}}.
\end{align*}


\paragraph{1st term} The two point function from this term is calculated as follows:
\begin{align*}
&\frac{\delta}{\delta J(P)} \frac{\delta}{\delta J(Q)}4 \bar \la \int {\cal D}\eta^\prime  Tr(\Delta \eta^\prime) Tr [ (\Delta J)^2 \Delta \eta^\prime]e^{-\int \frac{\eta'^2}{2}}\\
=& \frac{\delta}{\delta J(P)} \frac{\delta}{\delta J(Q)} 4 \bar \la \int {\cal D}\eta^\prime e^{-\int \frac{\eta'^2}{2}} \bigg\lbrace  \int_u \Delta (u-u)\eta^\prime(u) \bigg\rbrace \bigg\lbrace \int_{y,z} \Delta (x-y) J(y) \Delta(y-z)J(z)\Delta(z-x)\eta^\prime(x)\bigg\rbrace\\
=& 16 \bar \la \int {\cal D}\eta^\prime e^{-\int \frac{\eta'^2}{2}} \bigg\lbrace \int_u \Delta (u-u)\eta^\prime(u) \bigg\rbrace \lbrace  \Delta (x-P)  \Delta(P-Q)\Delta(Q-x)\eta^\prime(x)\rbrace\\
=&  16 \bar \la \Delta(0)\Delta(x-P)\Delta(P-Q)\Delta(Q-x).
\end{align*}

\paragraph*{3rd term}
\begin{align*}
&\frac{\delta}{\delta J(P)} \frac{\delta}{\delta J(Q)}\frac{8\bar \la}{N} \int {\cal D}\eta^\prime Tr [ \Delta \eta^\prime (\Delta J)^2 \Delta \eta^\prime ]e^{-\int \frac{\eta'^2}{2}}\\
=&\int {\cal D}\eta^\prime e^{-\int \frac{\eta'^2}{2}} \int_{y,z,u} \Delta(x-y) \eta^\prime(y) \Delta(y-z)J(z)\Delta(z-u)J(u)\Delta(u-x)\eta^\prime(x)\\
=& \frac{8\bar \la}{N} \Delta(0)\Delta(x-P)\Delta(P-Q)\Delta(Q-x).
\end{align*}
Because of the presence of $ \Delta (0)$, we can cancel these terms using a lower order counter-term.

\paragraph{2nd term:} $2 \bar{\lambda} \int \mathcal{D}\eta^\prime Tr(\Delta J \Delta \eta^\prime) Tr (\Delta J \Delta \eta^\prime) e^{-\frac{(\eta\prime)^2}{2}}$

\vspace{0.1 in}

To find $\langle \sigma \sigma \rangle$ ,
\begin{align*}
& \hf \frac{\delta}{\delta J(p)} \frac{\delta}{\delta J(Q)} \int \mathcal{D} \eta^\prime Tr(\Delta J \Delta \eta^\prime) Tr(\Delta J \Delta \eta^\prime) e^{-\frac{\eta^2}{2}}\\
= & \hf \frac{\delta}{\delta J(p)} \frac{\delta}{\delta J(Q)}\int \mathcal{D}\eta^\prime e^{-\int_x \frac{\eta^{\prime^2}}{2}}\int_{y}  \left[ \Delta(x-y)J(y) \Delta(y-x)\eta(x)\right] \int_{x_1,y_1}  \left[ \Delta(x_1-y_1)J(y_1) \Delta(y_1-x_1)\eta(x_1)\right]\\
=& \int \mathcal{D}\eta^\prime e^{-\frac{\eta^2}{2}}  \left[ \Delta(x-P) \Delta(P-x)\eta(x)\right] \int_{x_1}  \left[ \Delta(x_1-Q) \Delta(Q-x_1)\eta(x_1)\right] \\
=& \int_{x_1} \Delta(x-P) \Delta(P-x) \Delta(x_1-Q)\Delta(Q-x_1) \delta(x-x_1)\\
=&  \Delta(x-P) \Delta(P-x) \Delta(x-Q)\Delta(Q-x).
\end{align*}

In momentum space, the above expression translates to
\begin{align*}
&\int_{k,r} \frac{1}{(p+k)^2+m^2}\frac{1}{k^2+m^2}\frac{1}{(p+r)^2+m^2}\frac{1}{r^2+m^2}.
\end{align*}
We evaluate the above integral using Gaussian cutoff:
\begin{align}\label{int}
& \int_{k,r} \frac{e^{\frac{-(p+k)^2+m^2}{\Lambda^2}}}{(p+k)^2+m^2}\frac{e^{\frac{-(k)^2+m^2}{\Lambda^2}}}{(k)^2+m^2}\frac{e^{\frac{-(p+r)^2+m^2}{\Lambda^2}}}{(p+r)^2+m^2}\frac{e^{\frac{-(r)^2+m^2}{\Lambda^2}}}{(r)^2+m^2}\\
=\nonumber & \int_{k,r} \int_{x,y,u,v=\infty}^{\frac{1}{\Lambda^2}} e^{-\lbrace(p+k)^2+m^2 \rbrace x}e^{-\lbrace(k)^2+m^2 \rbrace y}e^{-\lbrace(p+r)^2+m^2 \rbrace u}e^{-\lbrace(r)^2+m^2 \rbrace v}.
\end{align}

Let's concentrate on integration of $x,y$ and $k$ first at $p=0$:
\begin{align*}
&\int_{x,y=\infty}^{\frac{1}{\Lambda^2}} \int \frac{d^4k}{(2\pi)^4} e^{-\lbrace(k)^2+m^2 \rbrace x} e^{- \lbrace k^2+m^2 \rbrace y }\\
& =\frac{1}{16\pi^2}\int_{x,y} \frac{1}{(x+y)^2}e^{-m^2(x+y)}\\
& = \frac{1}{16\pi^2}\int_x \int_{y=\infty}^{(x+\frac{1}{\Lambda^2})} \frac{1}{y^2} e^{-m^2 y}\\
&= \frac{1}{16\pi^2}\int_x \left[-\frac{e^{-m^2 (x+\frac{1}{\Lambda^2})}}{(x+\frac{1}{\Lambda^2})}+ m^2 Ei\big\lbrace-m^2\Big(x+\frac{1}{\Lambda^2}\Big) \Big\rbrace\right]\\
&= \frac{1}{16\pi^2}\int_{x=\infty}^{\frac{2m^2}{\Lambda^2}} \left[ -\frac{e^{-x}}{x}+m^2 Ei(-x) \right]\\
& = -\frac{1}{16\pi^2}Ei\Big(-\frac{2m^2}{\Lambda^2}\Big)+ \frac{m^2}{16\pi^2}\int_{x=\infty}^{\frac{2m^2}{\Lambda^2}} Ei(-x).
\end{align*}
The log term will come from first term only. We extract that:
\begin{align*}
=\frac{1}{16\pi^2}\big(-\log \frac{2m^2}{\Lambda^2}+\gamma_{EM}+ \dots\big)=\frac{1}{16\pi^2}\big( \log \Lambda^2+\gamma_{EM}+\dots \big),
\end{align*}
where
$\gamma_{EM}$ is the Euler-Mascheroni constant.

\paragraph{} Next, doing the integrals of $u,v$ and $r$, we get the same result. Combining both and extracting the log term, we get expression for \eqref{int} as 
\begin{align*}
\frac{1}{(16\pi^2)^2}(\log \Lambda^2+\gamma_{EM}+\dots)(\log \Lambda^2+\gamma_{EM}+\dots)= \frac{2}{(16\pi^2)^2}\gamma_{EM} \log \Lambda^2.
\end{align*}

So ,
\begin{align}\label{sigmab}
\langle \sigma \sigma \rangle_b =\frac{4 \bar \la}{(16\pi^2)^2} \gamma_{EM}\log \Lambda^2.
\end{align}

\paragraph{4th term} $\frac{4\bar \la}{N} \int \mathcal{D} \eta^\prime Tr [ \Delta J \Delta \eta^\prime \Delta J \Delta \eta^\prime ]e^{-\eta^{\prime^2}} $

\vspace{0.1 in}

To find $\langle \sigma \sigma \rangle $,
\begin{align*}
& \hf \frac{\delta}{\delta J(p)} \frac{\delta}{\delta J(Q)} \int \mathcal{D} \eta^\prime Tr[(\Delta J \Delta \eta^\prime)(\Delta J \Delta \eta^\prime)] e^{-\frac{\eta^2}{2}}\\
= & \hf  \frac{\delta}{\delta J(p)} \frac{\delta}{\delta J(Q)}\int \mathcal{D}\eta^\prime \int_{y,z,u} e^{-\int_x \frac{\eta^{\prime^2}}{2}} \times  \left[ \Delta(x-y)J(y) \Delta(y-z)\eta(z) \Delta (z-u) J(u) \Delta(u-x) \eta(x) \right].
\end{align*}

Proceeding like before, we get
\begin{align*}
\Delta (P-x)\Delta(x-P)\Delta(x-Q)\Delta(Q-x).
\end{align*}
So, we get the same Feynman diagram as for the 3rd term. The final value is
\begin{align}\label{sigmac}
\langle \sigma \sigma \rangle_c= \frac{8 \bar \la}{N} \frac{1}{(16\pi^2)^2} \gamma_{EM}\log \Lambda^2.
\end{align}

We get the two point function as
\begin{align*}
\langle \sigma \sigma \rangle= \gamma_{EM} \log \Lambda^2 \frac{\bar \la}{(16\pi^2)^2}. (4+\frac{8}{N})
\end{align*}
Replace $\log \Lambda^2= 2t $ and $\frac{\la}{4!}= \frac{\bar \la}{2N}$ to get
\begin{align*}
\langle \sigma \sigma \rangle= & \gamma_{EM} 2t \frac{1}{(16\pi^2)^2} \frac{2N \la}{4!} (4+\frac{8}{N})\\
=& (\gamma_{EM} t) \frac{2}{3} \frac{\la}{(16\pi^2)^2}(N+2).
\end{align*}

Hence, the anomalous dimension of $\phi^2$ is  given by
\begin{center}

\boxed{ \gamma_m= \frac{\la}{16\pi^2} \frac{1}{3}(N+2)}.

\end{center}


%% file: RegERGAppendix_input.tex
\section{ERG equation for Auxiliary fields} \label{ERGeq}

In this Appendix we give details of the derivation of the ERG equation for the scalar composite auxiliary field action in the free scalar theory discussed in Section 3. It is also shown that the action is a fxed point of the ERG equation.

\subsection{Equation for $S_\Lambda[\phi_l,J]$} \label{actionJ}

We verify that \eqref{2.29} satisfies \eqref{polchERG}.

We have
\begin{align*}
	\frac{\dd S_{\lm,I}[\phi_l,J]}{\dd \phi_l(x)}=&-2i\int_u J\phi_l(u) \frac{1}{(1-2i\DD_h J)}_{ux}\\
	\implies \frac{\dd^2S_{\lm,I}[\phi_l,J]}{\dd \phi_l(x)\dd \phi_l(y)}\bigg|_{\phi_l=0}=&-
	2iJ(x)\frac{1}{(1-2i\DD_h J)}_{xy}.
\end{align*}
Since we are going to set $\phi_l=0$ at the end, we only need this term. So, the RHS of the Polchinski's equation is
\be \label{polRHS}
\hf\int_x\int_y\dot \DD_{hxy} \frac{\dd^2S_{\lm,I}[\phi_l,J]}{\dd \phi_l(x)\dd \phi_l(y)}\bigg|_{\phi_l=0}=
-\hf\int_x\int_y \bigg(\frac {2i\dot \DD_h J}{1-2i \DD_h J}\bigg)_{xy}
\ee
\[
=-\hf\bigg[\int_x 2iJ(x)\dot \DD_h(0)+
\int_x\int_y 2iJ(x)\dot \DD_h(x-y) 2iJ(y)\DD_h(x-y)+....\bigg],
\]
which is just a sum of one loop graphs with $J$'s attached and one propagator differentiated. It is also clear that
\be  \label{2.31}
-\hf\int_x\int_y \bigg(\frac {2i\dot \DD_h J}{1-2i \DD_h J}\bigg)_{xy}=\frac{\p}{\p t} Tr \ln [1-2i\DD_h J]_{xy},
\ee
and
\begin{align}
	&\hf\int_x\int_y\dot \DD_{hxy}\bigg(\frac{\dd S_{\lm,I}[\phi_l,J]}{\dd \phi_l(x)}\bigg)\bigg(\frac{\dd S_{\lm,I}[\phi_l,J]}{\dd \phi_l(y)}\bigg)\nonumber\\
	=&\hf\int_x\int_y\dot \DD_{hxy}
	\bigg(2i\int_u J\phi_l(u) \frac{1}{(1-2i\DD_h J)}_{ux}\bigg)\bigg(2i\int_v J\phi_l(v) \frac{1}{(1-2i\DD_h J)}_{vy}\bigg)\nonumber\\
	=&\label{2.32}
	\hf\int_x\int_y
	\bigg(2i\int_u J\phi_l(u) \frac{1}{(1-2i\DD_h J)}_{ux}\bigg)\dot \DD_{hxy}\bigg(2i\int_v  \frac{1}{(1-2iJ\DD_h )}_{yv}J\phi_l(v)\bigg)\\
	=& \label{2.33}
	\frac{\p}{\p t} \int_x\int_y iJ\phi_l(x)\bigg(\frac{1}{1-2i\DD_hJ}\bigg)_{xy} \phi_l(y).
\end{align}
Combining  equations \eqref{2.29} and \eqref{polRHS}-\eqref{2.33}, we see
\be 
\hf\int_x\int_y\dot \DD_{hxy} \bigg(\frac{\dd^2S_{\lm,I}[\phi_l,J]}{\dd \phi_l(x)\dd \phi_l(y)}
-\bigg(\frac{\dd S_{\lm,I}[\phi_l,J]}{\dd \phi_l(x)}\bigg)\bigg(\frac{\dd S_{\lm,I}[\phi_l,J]}{\dd \phi_l(y)}\bigg)\bigg)= \frac{\p}{\p t}S_{\lm,I}[\phi_l,J].
\ee
Thus, we see that, as expected, the action obeys the ERG equation. 

Although we had started with a free theory, the ERG equation for the auxiliary field picks up potential terms. Below, we confirm that it is indeed a fixed point.

\paragraph{Fixed Point Condition}
To see the fixed point nature, one has to write the action in terms of dimensionless variables. This changes the time derivative:
\[
\frac{d}{dt}=\frac{\p}{\p t}\bigg|_{dimensionless} -G_{dil},
\]
where $G_{dil}$ is the contribution to the time derivative due to the powers of $\lm$ that are there simply for dimensional reasons, i.e., due to the engineering dimensions of the variables, the fields and momenta. 

\[
G_{dil}= -N_J[J] -N_\phi[\phi]-N_p.
\]
$[J]$ is the dimension of the field $J$, $N_J$ is the number of fields in any term. $N_p$ is the number of factors of momenta in the coefficient functions in any term. Formally,
\[
N_p= \sum_ip_i\frac{\p }{\p p_i}.
\]  
At a fixed point, one expects
\be   \label{2.36}
\frac{\p}{\p t}\bigg|_{dimensionless}=0.
\ee
The physical meaning is that, at the fixed point, when the action is written in terms of  dimensionless variables with $\lm$ setting the scale, all $\lm$ dependence disappears, i.e., the dimensionless couplings do not ``run".

Consider a typical term:
\[
\int_{p_1}...\int_{p_n}\dd(p_1+...+p_n)
\int_k \DD_h(p_1+k)\DD_h(p_1+p_2+k)+...+\DD_h(p_1+...+p_n+k)J(p_1)...J(p_n)
\]
\[
\equiv\int_{p_1}...\int_{p_n}\dd(p_1+...+p_n)
{\cal F}_n(p_1,...,p_n)J(p_1)...J(p_n).
\]
$N_p$ gives $-D$ acting on the momentum conserving delta function. 
Using 
\[
p\frac{\p }{\p p}\DD_h(p) = -2 \DD_h(p) -2K'(p^2),
\]
one finds
\be  \label{31}
\sum_ip_i\frac{\p }{\p p_i}{\cal F}_n(p_1,...,p_n)=(D-2n){\cal F}_n(p_1,...,p_n)
+ \frac{d}{dt}{\cal F}_n(p_1,...,p_n).
\ee
We have used 
\[
\dot \DD_h = -\frac{d}{dt}\frac{K(p^2/\lm^2)}{p^2}=
 +2\lm^2 \frac{d}{d\lm^2} \frac{K(p^2/\lm^2)}{p^2}=-2 \frac{d}{d\bp^2}K(\bp^2).
\]
The $D$ in \eqref{31} cancels the $-D$ of momentum delta function. $n=N_J$ and $[J]=2$.
Thus,
\[
G_{dil}=-N_J([J]-2) -\frac{d}{dt}=-\frac{d}{dt}
\]
for the term containing only $J$. The other terms have an extra factor of $\int_x J(x)\phi^2$ which is also dimensionless and do not contribute to $G_{dil}$. Thus, the entire time derivative comes from $G_{dil}$, and thus,
\eqref{2.36} is satisfied, and we have a fixed point action. Of course, since the underlying theory is free, this is expected.

\subsection{Equation for $S_\Lambda[\si]$} \label{actionsigma}

As argued in main section, we consider N-scalar field theory. Our starting point is \eqref{2.51}, i.e.,
\begin{align}
	e^{-S_{\lm,I}[\phi_l,\si]}&=
	\int {\cal D}\chi ~e^{i\int \chi \si -i\int \chi \phi_l^2}e^{ -\frac{N}{2} Tr \ln [\frac{\sqrt N }{\DD_h}+2i\chi]^{-1} }e^{-\hf \int_x\int_y(2\chi \phi_l^I(x)([\frac{\sqrt N }{\DD_h}+2i\chi]^{-1})_{xy}(2\chi \phi_l^I(y))}\nonumber\\
	&=\int {\cal D}\chi ~e^{i\int \chi \si -\frac{N}{2} Tr \ln [\frac{\sqrt N }{\DD_h}+2i\chi]^{-1}-i\int_x\int_y \chi \phi_l^I(x) ([1+\frac{2i\DD_h\chi}{\sqrt N}]^{-1})_{xy}\phi_l^I(y) }
\end{align}

Expanding the logarithm produces a linear term in $\chi$ which is $\int_x i\sqrt N \DD_h(0) \chi(x)$. As in \eqref{2.51}, the Fourier transform is w.r.t. $\si -\sqrt N\DD_h(0)=\si'$. This will be important below.

Thus, we let 
\be
e^{-S_{\lm,I}'[\phi_l^I,\si']} \equiv \int {\cal D}\chi ~e^{i\int \chi \si'-S_{\lm,I}[\phi_l^I,\chi]},
\ee
where, in $S_{\lm,I}[\phi_l^I,\chi]$, the linear term in $\chi$ has been subtracted. We put $\phi_l=0$ to get $S_\Lambda[\si]$.

\vspace{0.1 in}

\eqref{2.40} is modified to include $N$ dependence to
\be \label{2.44}
\frac{\p}{\p t}e^{-S_{\lm,I}[0,\si]}=
-\frac{N}{2}\int_x\int_y \dot \DD_h(x-y)\Bigg(\frac {\frac{2}{\sqrt N} \frac{1}{i}\frac{\dd}{\dd \si}}{1+\frac{1}{i}\frac{2 \DD_h}{\sqrt N} \frac{\dd }{\dd \si}}\Bigg)_{xy} 
e^{-S_{\lm,I}[0,\si]}.
\ee
Let us expand this:
\begin{align}
	\frac{\p}{\p t}e^{-S_{\lm,I}[0,\si]}=&
	-\Big(\frac{1}{i}\frac{N}{2}\int_x 
	\frac{2\dot \DD_h(0)}{\sqrt N}
	\frac{\dd}{\dd \si(x)}
	+\frac{N}{2}\int_x\int_y\frac{2\dot \DD_h(x-y)}{\sqrt N}\frac{2 \DD_h(x-y)}{\sqrt N}\frac{\dd^2}{\dd \si(x)\dd\si(y)}\nonumber\\
	&+ \frac{1}{i}\frac{4}{\sqrt N}\int_{x,y,z}\dot \DD_h(x-y)\DD_h(x-z)\DD_h(z-y)\frac{\dd^3}{\dd \si(x)\dd \si(y)\dd\si(z)}+....\Big)e^{-S_{\lm,I}[0,\si]}.\label{2.45}
\end{align}
\vspace{0.2 in}

\paragraph{ Linear Term:}

The linear term is unpleasant. Its origin is in the tadpole diagram or self-contraction of
the $\phi$'s in $\phi^2$:
\be
\phi^2(x)= :\phi^2(x):+ \frac{1}{i}\sqrt N \DD_h(0).
\ee
The $:...:$ stands for normal ordering. Below, we show that by redefining $\si$, one can get rid of the linear term in \eqref{2.45}. Let us write
\be
\si = \si'+\frac{1}{i}\sqrt N\DD_h(0).
\ee
Since $\si$ has no $t$ dependence, we have
\be
\frac{d\si'}{dt}=-\frac{1}{i}\sqrt N\dot\DD_h(0).
\ee
We have also defined earlier
\be
S_\lm[\si]=S_\lm[\si'+\sqrt N\DD_h(0)]=S'_\lm[\si'].
\ee
We see that, in addition to the explicit dependence on $\lm$ in $S_\lm'[\si']$, there is an implicit $\lm$ dependence inside $\si'$.
Thus,
\be   \label{2.49}
\frac{d S_\lm[\si]}{d t}= \frac{d S_\lm'[\si']}{d t}=\frac{\p S_\lm'[\si']}{\p t}-\frac{1}{i}
\sqrt N\dot \DD_h(0)\int_x\frac{\dd S'[\si']}{\dd \si'(x)}.
\ee

Substituting \eqref{2.49} in the LHS of \eqref{2.45}, we see that the linear derivative term cancels out, and we are left with a Polchinski type ERG equation.
\begin{align}
	\frac{\p}{\p t}e^{-S_{\lm,I}[\si]}=&
	\Big(
	-\int_x\int_y \frac{d (\DD_h(x-y))^2}{dt}\frac{\dd^2}{\dd \si(x)\dd\si(y)}\\
	&+\frac{1}{i}\frac{4}{\sqrt N}\int_{x,y,z}\dot \DD_h(x-y)\DD_h(x-z)\DD_h(z-y)\frac{\dd^3}{\dd \si(x)\dd \si(y)\dd\si(z)}+....\Big)e^{-S_{\lm,I}[\si]}.\nonumber
\end{align}

Hereafter, we work with $\si'$ and $S'[\si']$, drop the linear term in \eqref{2.45}, and also drop the primes for notational simplicity.
%

\vspace{0.2 in}

\paragraph{Quadratic Term:}

The second term is of the type in  Polchinski ERG equation. Its coefficient is the time derivative of the $\si$ field propagator and is of $O(1)$ thanks to our choice of normalization for the $\phi$ kinetic term. The propagator for $\si$ is the $\phi^2$ two point correlation, which is proportional to $\DD_h(x-y)^2$. If $\si$ had been an elementary field (like $\phi$) the equation would have ended there. But, $\si$ stands for a composite field and one expects Polchinski equation to be modified. The higher order correction terms are proportional to powers of $1/\sqrt N$. 

\vspace{0.2 in}

\paragraph{ Cubic Term:} The cubic term is also unpleasant.  However, one can evaluate this to leading order in $1/N$ by acting on the $O(1)$ of $S_\lm[\si]$,
which can be obtained as follows.
The quadratic $\chi$ term in $S_{\lm,I}[0,\chi]$ is the $O(1)$ kinetic term
\[
- \int_x\int_y \chi(x) \DD_h(x-y)^2\chi(y).
\]
Doing the Gaussian $\chi$ integral gives for the quadratic $\si$ the $O(1)$ term 
\be
-\frac{1}{4}\int_x\int_y \si(x) \DD_h^{-2}(x-y)\si(y).
\ee
Acting with the cubic in derivatives term, we obtain
\[
-\frac{4}{\sqrt N}. \frac{1}{8}\int_{x,y,z}\int_{u,v,w}
\dot \DD_h(u-v)\DD_h(u-w)\DD_h(w-v)\frac{1}{\DD^2(u-x)}
\frac{1}{\DD^2(v-y)}\frac{1}{\DD^2(w-z)}\si(x)\si(y)\si(z)
\]
\be
=\frac{1}{2\sqrt N}\int_{x,y,z}f(x,y,z)\si(x)\si(y)\si(z),
\ee
where
\be
f(x,y,z)\equiv \int_{u,v,w}
\dot \DD_h(u-v)\DD_h(u-w)\DD_h(w-v)\frac{1}{\DD^2(u-x)}
\frac{1}{\DD^2(v-y)}\frac{1}{\DD^2(w-z)}.
\ee
This is a modification of the ERG equation---if the leading term of Polchinski's ERG equation is like a free particle Hamiltonian, this one  is like a potential term in a Hamiltonian.


%% file: ERGVector_input.tex
\section{ERG Equation Derivation for the Vector Composite}\label{vectorapp}

In this Appendix, we give some details of the ERG equation calculation for the vector case discussed in Section \ref{polch}.

The following equation is valid for Wilson action interaction part with the auxiliary field taken to be an external field.
\be   \label{32}
\frac{\p S_{\lm,I} }{\p t}\bigg|_{\phi_l=0}=
\hf \frac{1}{\sqrt N} \int_x\int_y \dot \DD _{hxy}\bigg[\frac{\dd^2 S_{\lm,I}}{\dd \phi_l ^I(x)\dd \phi_l ^I(y)}-\frac{\dd S_{\lm,I}}{\dd \phi_l ^I(x)} \frac{\dd S_{\lm,I}}{\dd \phi_l ^I(y)}\bigg]\bigg|_{\phi_l=0}.
\ee

\subsection{Evaluation of $S_\lm[\phi_l^I,\sigma_\mu^{IJ}]$}

\begin{enumerate}
%

\item Write
\[
\phi^I=\phi_l^I+\phi_h^I~;~~\Delta=\DD _l+\DD_h;
\]
\[
 \phi^I\DD^{-1}\phi^I=\phi_h^I\DD_h^{-1}\phi_h^I+\phi_l^I\DD_l^{-1}\phi_l^I.
\]
Let
\begin{align}
	S_{2B} =&\int \hf \sqrt N \phi_h^I\DD_h^{-1}\phi_h^I -\chi(\phi_h^I \phi_h^I + 2 \phi_l\phi_h)  -\sqrt N \chi^\mu_{AB} (\phi_h^A \olra {\p_\mu} \phi_h^B +2 \phi_l^A \olra{\p_\mu} \phi_h^B)\nonumber\\
	&-\hf \sqrt N \chi^\mu_{AC}\chi_{\mu AD}[\phi_h^C \phi_h^D+ 2\phi_l^C \phi_h^D].\label{S2B}
\end{align}
What is left is
\begin{align}
	S_{1B}=&\int \hf \sqrt N \phi_l^I\DD_l^{-1}\phi_l^I-\chi\phi_l^I \phi_l^I - \frac{1}{2\bar u}\Big(\chi+{r\sqrt N\over 2}\Big)^2 + \sqrt N\chi^\mu_{AB}(\si_\mu^{AB}-\phi_l^A \olra {\p_\mu} \phi_l^B)\nonumber\\ 
	&-\hf \sqrt N \chi^\mu_{AC}\chi_{\mu AD}\phi_l^C\phi_l^D.\label{S1B}
\end{align}

Let us separate the terms quadratic and linear in $\phi_h$.
\item {\bf Quadratic Terms}

We turn to the quadratic terms. These are of the same form as in Section \ref{vector} with the replacement $\DD\to \DD_h$.
\[
S_{2B}^{hh}= \hf \sqrt N \phi_h^I\DD_h^{-1}\phi_h^I -\chi \phi_h^I\phi_h^I
-\sqrt N\chi^\mu_{IJ}\phi_h^I\olra {\p_\mu} \phi_h^J-\hf \sqrt N \chi^\mu_{AC}\chi_{\mu AD}\phi_h^C \phi_h^D.
\]
 Some notation:
Definition of $\DD_h^{-1}$:
\[
\int _u \DD_h ^{-1}(x,u)\DD_h(u,y)=\dd(x-y).
\]
Thus, if $\DD_h^{-1}(x,y)= \p_x^2\dd(x-y)$ then $\DD_h(x,y)=\frac{1}{4\pi^2(x-y)^2}$ in 4 dimensions.

\[
\int_x \chi(x)\phi_h^I(x)\phi_h^I(x)=\int_x\int_y \phi_h^I(x)\chi(x)\dd(x-y)\phi_h^I(y);
\]
\begin{align*}
	&\sqrt N\int_x \chi^\mu_{IJ}(x)\phi_h^I(x)\olra{\p_\mu} \phi_h^J(x)\\
	=&
	\sqrt N\int_x\int_y~\chi^\mu_{IJ}(x)[\phi_h^I(x)\p_{x^\mu} \dd(x-y)\phi_h^I(y)-\p_{x^\mu}\phi_h^I(x) \dd(x-y)\phi_h^J(y)]\\
	=&	\sqrt N\int_x\int_y~\phi_h^I(x)[2\chi^\mu_{IJ}(x)\p_{x^\mu} \dd(x-y)+\p_{x^\mu}\chi^\mu_{IJ}(x) \dd(x-y)]\phi_h^J(y).
\end{align*}
Thus, we can write $S_{2B}^{hh}$ as
\[
\hf\int_x\int_y~\phi_h^I(x)[\underbrace{ \sqrt N \DD_h^{-1}(x-y)}_{A}\dd_{IJ}
\]
\[-\underbrace{(2\chi(x)\delta_{IJ}+\sqrt N (4\chi^\mu_{IJ}(x)\p_{x^\mu} 
	+ 2\p_{x^\mu}\chi^\mu_{IJ}(x)+ \chi^\mu_{KI}\chi_{\mu}^{KJ})) \dd(x-y)}_{B}]\phi_h^J(y)
\]
\be   \label{25}
\equiv \hf\int_x\int_y~\phi_h^I(x)O^{hh}_{IJ}(x,y)\phi_h^J(y).
\ee
The $\phi_h$ integral gives $(Det ~O_{IJ}^{hh})^{-\hf}$. We then write $O_{IJ}^{hh}= A\dd_{IJ}-B= A(I- A^{-1}B)$. Then,
\[
Det [A(I-A^{-1}B)]^{-\hf}=[Det A]^{-\hf} [Det (I-A^{-1}B)]^{-\hf}.
\]
Here, $A=  \sqrt N\DD^{-1}_{h}$ is field independent, so, we can ignore its determinant. That leaves
\[
Det[I-  \frac{1}{\sqrt N}\DD_h B]^{-\hf}= e^{-\hf Tr \ln [I-\frac{1}{\sqrt N}\DD_h B]}.
\]
Finally, $\ln (1-x)= -(x+\hf x^2 +\frac{1}{3}x^3+...)$, so, we get
\be  \label{26}
Tr \ln [I-\frac{1}{\sqrt N}\DD_h B]= -(Tr\bigg(\frac{1}{\sqrt N}\DD_h B\bigg)+\hf Tr \bigg(\frac{1}{\sqrt N}\DD_h B\bigg)^2 +...),
\ee
where
\[  
B=2\chi(x)\dd(x-y)\dd_{IJ}+\sqrt N 4\chi^\mu_{IJ}(x)\p_{x^\mu} \dd(x-y)+\sqrt N 2\p_{x^\mu}\chi^\mu_{IJ}(x) \dd(x-y)
\]
\be \label{eqnB}
+\hf \sqrt N \chi^\mu_{KI}(x)\chi_{\mu KJ}(x)\dd(x-y)=J_{IJ}\dd(x-y),
\ee
where we have, in anticipation, defined
\be    \label{jij}
 2\chi (u)\dd_{IJ} + \sqrt N (2 \p_\mu\chi^\mu_{IJ}(u)+ 4\chi^\mu_{IJ}(u)\frac{\p}{\p u^\mu}+ \p_{x^\mu}\chi^\mu_{IJ}(x)+ \chi^\mu_{KI}(x)\chi_{\mu KJ}(x))\equiv
 J_{IJ}(u).
\ee

The linear term in \eqref{26} corresponds to tadpole graphs. The $\chi$ tadpole has been discussed---the value of $r$ can be tuned to $r_c$ to cancel this. The $\chi^\mu$ tadpoles must be proportional to the momentum of the $\chi^\mu$ field and must vanish in the vacuum.

\item {\bf Linear Terms in $S_{2B}$}
\be
S_{2B}^h=-\int_x 2\chi \phi_l^I\phi_h^I +\sqrt N 2\chi^\mu_{IJ}\phi_l^I \olra{\p_\mu}\phi_h^J+ \sqrt N \chi^\mu_{KI}\chi_{\mu KJ}\phi_l^I \phi_h^J.
\ee
This can be written after an integration by parts as
\[
S_{2B}^h=-\int_x (2\chi \phi_l^I\dd_{IJ} +\sqrt N 2 \p_\mu\chi^\mu_{IJ}\phi_l^I+\sqrt N 4\chi^\mu_{IJ}\p_\mu\phi_l^I+ \sqrt N \chi^\mu_{KI}\chi_{\mu KJ}\phi_l^I)\phi_h^J
\]
\be   
\equiv -\int_x~j_J(x)\phi_h^J(x),
\ee
with
\be    \label{27}
j_J(x)= 2\chi(x)\phi_l^I\dd_{IJ} + \sqrt N 2 \p_\mu\chi^\mu_{JI}\phi_l^I+ \sqrt N 4\chi^\mu_{JI}\p_\mu\phi_l^I+ \sqrt N \chi^\mu_{KI}\chi_{\mu KJ}\phi_l^I,
\ee
(using the antisymmetry of $\chi^\mu_{IJ}=-\chi^\mu_{JI}$).
\item

Thus, combining \eqref{25} and \eqref{27}, we get
\[
\int {\cal D}\phi_h^I e^{-S_{2B}}=\int {\cal D}\phi_h^I~e^{-\hf\int_x\int_y~\phi_h^I(x)O^{hh}_{IJ}(x,y)\phi_h^J(y) +\int_x~j_J(x)\phi_h^J(x)}
\]
\be
=
e^{-\hf Tr \ln [I-\frac{1}{\sqrt N}\DD_h B] +\hf \int_x\int_y j_I(x) [O_{IJ}^{hh}(x,y)]^{-1}j^J(y) }\equiv e^{-S_{2,\lm}[\phi_l^I,\chi,\chi^\mu_{IJ}]}.
\ee
Here, one can write
\[
[O_{IJ}^{hh}(x,y)]^{-1}=(A-B)^{-1}= (A(I-A^{-1}B))^{-1}=(I-A^{-1}B)^{-1}A^{-1}\]
\[= A^{-1}+A^{-1}BA^{-1}+A^{-1}BA^{-1}BA^{-1}+...
\]

Thus,
\begin{align}
	&\hf \int_x\int_y j_I(x) [O_{IJ}^{hh}(x,y)]^{-1}]j^J(y)\nonumber\\
	&=\hf \int_x\int_y j_I(x) \Big[\frac{1}{\sqrt N}\DD_h(x,y) +\frac{1}{\sqrt N}\DD_h B \frac{1}{\sqrt N}\DD_h+...\Big] j^I(y),\label{30}
\end{align}
where $B$ is given in \eqref{eqnB}.

Thus, our starting point \eqref{15} can be written as
\begin{align}
	e^{-S_\lm[\phi_l^I,\sigma^\mu_{IJ}]}&=\int {\cal D}\phi_h^I {\cal D} \chi {\cal D} \chi^\mu_{AB}~e^{-S_B[\phi_l^I,\chi,\chi^\mu_{AB},\sigma_\mu^{AB},\phi_h^I]}\nonumber\\
	&=
	\int {\cal D} \chi {\cal D} \chi^\mu_{AB}~e^{-S_{1B}[\phi_l^I,\chi,\chi^\mu_{AB}]-S_{2,\lm}[\chi,\chi^\mu_{AB},\phi_l^I]+i\int _x \si_\mu^{AB}\chi^\mu_{AB} }.
\end{align}
 $\si_\mu^{AB}$ acts as a source for $\chi^\mu_{AB}$. Removing the kinetic term $\sqrt N\int \hf \phi_l^I \DD_l^{-1}\phi_l^I$ from $S_{\lm}[\phi_l^I,\si_\mu^{AB}]$ gives $S_{\lm,I}[\phi_l^I,\si_\mu^{AB}]$ which obeys \eqref{20}.  The equation will be evaluated for $\phi_l=0$.  This is only for simplicity---we are not interested in the information contained in the $\phi_l$ dependence.
Thus, we will denote $S_{\lm,I}[0,\si_\mu]=S_\lm[\si_\mu]$ below, where $\phi_l$  has been set to zero.

\end{enumerate} 

%


%

\subsection{Evaluation of $\frac{\dd S_{\lm,I}}{\dd \phi_l ^I(x)}$,$\frac{\dd ^2 S_{\lm,I}}{\dd \phi_l ^I(x)\dd \phi_l ^I(y)}$}

Let us evaluate the functional derivatives:
\begin{enumerate}
\item
\[
\frac{\dd S_{\lm,I}}{\dd \phi_l ^I(x)}=\frac{\dd S_{1,I}}{\dd \phi_l ^I(x)}+\frac{\dd S_{2,\lm}}{\dd \phi_l ^I(x)}
\]
Here $S_{1,I}$ is  $S_1$ minus the kinetic term. (See eqn \eqref{S1B} and \eqref{S2B}.)

Then,
\[
S_{1,I}=\int -\chi\phi_l^I \phi_l^I - \frac{1}{2\bar u}\bigg(\chi+{r\sqrt N\over 2}\bigg)^2 + \chi^\mu_{IJ}\bigg(\si_\mu^{IJ}-\sqrt N\phi_l^I \olra {\p_\mu} \phi_l^J\bigg)
\]
\be
-\hf \sqrt N \chi^\mu_{KI}\chi_{\mu KJ}\phi_l^I\phi_l^J.
\ee
Therefore,
\[   
\frac{\dd S_{1,I}}{\dd \phi_l ^I(x)}= -(2 \chi \dd_{IJ}\phi_l^J(x) + \sqrt N 4 \chi^\mu _{IJ}\p_\mu \phi^J_l + \sqrt N 2 \p_\mu \chi^\mu _{IJ} \phi_l^J + \sqrt N \chi^\mu_{KI}\chi_{\mu KJ}\phi_l^J)
\]
\be    \label{33}
=-j_I(x)=-J_{IJ}\phi_l^J .
\ee
Also,
\be
\frac{\dd S_{2,\lm}}{\dd \phi_l ^I(x)}= \int_u \frac{\dd S_{2,\lm}}{\dd j^J(u) }\frac{\dd j^J(u)}{\dd \phi_l ^I(x)}.
\ee

From \eqref{27},
\[
j_J(x)= 2\chi(x)\dd_{JI}\phi_l^I + \sqrt N 2 \p_\mu\chi^\mu_{JI}\phi_l^I+ \sqrt N 4\chi^\mu_{JI}\p_\mu\phi_l^I+ \sqrt N \chi^\mu_{KI}\chi_{\mu KJ}\phi_l^I,
\]
and the relevant part of $S_{2,\lm}$ is, (from \eqref{30}),
\[
\hf \int_x\int_y j_I(x) [O_{IJ}^{hh}(x,y)]^{-1}]j^J(y)=\frac1{2\sqrt N} \int_x\int_y j_I(x) [\DD_h(x,y) +\DD_h B \frac{1}{\sqrt N}\DD_h+...] j^I(y).
\] 
Thus, the leading term is
\[\frac{\dd S_{2,\lm}}{\dd j^J(u) }=\int_v \frac{1}{\sqrt N}\DD_h(u,v)j^J(v) \equiv \frac{1}{\sqrt N}(\DD_hj^J)(u),
\]
and
\be    \label{36}
\frac{\dd j_J(u)}{\dd \phi_l ^I(x)}= (2\chi (u)\dd_{JI} +\sqrt N (2 \p_\mu\chi^\mu_{JI}(u)+  4\chi^\mu_{JI}(u)\frac{\p}{\p u^\mu}+ \chi^\mu_{KI}\chi_{\mu KJ}(u)))\dd(u-x)
\ee
\[\equiv
J_{JI}(u)\dd(u-x).
\]
Thus,
\[
\frac{\dd S_{2,\lm}}{\dd \phi_l ^I(x)}=\frac{1}{\sqrt N}\int_u
(\DD_hj^J)(u)J_{JI}(u)\dd(u-x)
\]
\be   \label{38}
=\frac{1}{\sqrt N}J_{IJ}(x)(\DD_h j^J)(x).
\ee
Combining \eqref{38} and \eqref{33}, we get
\be   \label{39}
\frac{\dd S_{\lm,I}}{\dd \phi_l ^I(x)}=\frac{1}{\sqrt N}J_{IJ}(x)(\DD_hj^J)(x) -J_{IJ}(x)\phi_l^J(x).
\ee
This expression is linear in $\phi_l^I$. So, the second term in the RHS of the ERG equation  \eqref{32} vanishes and we are left with the first term. Accordingly, we need 
$\frac{\dd^2 S_{\lm,I}}{\dd \phi_l ^I(x)\dd \phi_l ^I(y)}$.
\[
\frac{\dd (\DD_hj^J)(x)}{\dd \phi_l^I(y)}=\frac{\dd}{\dd \phi_l^I(y)}\int_u \DD_h(x-u)j^J(u)
\]
\[=
\int_u \DD_h(x-u)\frac{\dd j^J(u)}{\dd \phi_l^I(y)}=\int_u \DD_h(x-u)J_{JI}(u)\dd(u-y)
\]
\[=\int_u \DD_h(x-u)(2\chi (u)\dd_{JI} + \sqrt N 2 \p_\mu\chi^\mu_{JI}(u)- \sqrt N 4\chi^\mu_{JI}(u)\frac{\p}{\p y^\mu}+ \sqrt N \chi^\mu_{KI}\chi_{\mu KJ}(u))\dd(u-y)
\]
\[=
\DD_h(x-y)(2\chi (y)\dd_{IJ} + \sqrt N 2 \p_\mu\chi^\mu_{IJ}(y)+\sqrt N \chi^\mu_{KI}\chi_{\mu KJ}(y))+ \sqrt N 4\chi^\mu_{IJ}(y)\p_{y^\mu}\DD_h(x-y).
\]
Thus,
\be   \label{40}
\frac{\dd (\DD_hj^J)(x)}{\dd \phi_l^I(y)}=
J_{IJ}(y)\DD_h(x-y).
\ee
(Note the interchange of $I,J$ in the last step. This is because of the two minus signs relative to $J_{JI}$.)
  
Combining \eqref{38} and \eqref{40}, we get
\[
\frac{\dd ^2  S_{2,\lm}}{\dd \phi_l ^I(x)\dd \phi_l ^I(y)}=\frac{1}{\sqrt N}J_{IJ}(x)\frac{\dd (\DD_h j^J)(x)}{\dd \phi_l^I(y)}=\frac{1}{\sqrt N}J_{IJ}(x)J_{IJ}(y)\DD_h(x-y).
\]
This is the contribution from the first term in \eqref{30}.
Thus, finally, from \eqref{39},
\be
\boxed{\frac{\dd ^2 S_{\lm,I}}{\dd \phi_l ^I(x)\dd \phi_l ^I(y)}=\frac{1}{\sqrt N}J_{IJ}(x)J_{IJ}(y)\DD_h(x-y)- J_{JI}(x)\dd(x-y)\dd_{IJ}}.
\ee
\end{enumerate}
This is the leading term in the ERG equation for the vector auxiliary action used in Section \ref{polch}.


%% file: f-constraints_input.tex
\section{Constraints on $f$ and the analyticity of the boundary term} \label{f-c}

In this appendix, we verify that the field redefinition we have done to obtain the AdS kinetic term is not illegal, i.e.,  the function $1/f$ involved in the redefinition has to be analytic everywhere. Since the function is expressed in terms of Bessel functions, we need only check for analyticity at the boundary, as $z\to 0$. In \eqref{boundaryterm}, we are left with a boundary term after redefining the field.   We show here that the integrand of the boundary term in the action is analytic. As long as the term is analytic in $p$, we can remove it from the action by adding counter-terms.

As written in Section \ref{map} and \cite{Sathiapalan:2017},
\begin{equation}
	1/f=A(p) z^{\Dt} K_\nu (pz) + B(p) z^{\Dt} I_\nu(pz),
\end{equation}
with $\nu^2= m^2 +\frac{D^2}{4}$, and $A(p)$ and $B(p)$ being arbitrary functions that we can tune to make $1/f$ analytic at $z\to0$, and to get the desired boundary conditions for the Green function. In \cite{Sathiapalan:2020}, these were chosen to be $A(p)=p^\nu$ and $B(p)=-\frac1{\gamma}p^\nu$ for a positive $\nu$. To ensure that Green function boundary behaviour is as desired, we just need to have $A(p)B(p)\sim p^{-2|\nu|}$.
The modified Bessel functions $I_\nu(x)$ and $K_\nu(x)$ have the following expansions in $x$.
\begin{align}
	I_\nu(x)=&\sum_{k=0}^{\infty}\frac{1}{\Gamma(k+\nu+1)k!}\Big(\frac x2\Big)^{2k+\nu},\\
	K_\nu(x)=&\hf \Big[\Gamma(\nu)\Big(\frac x2\Big)^{-\nu}\Big(1+\frac{x^2}{4(1-\nu)}+\ldots\Big) +\Gamma(-\nu)\Big(\frac x2\Big)^{\nu}\Big(1+\frac{x^2}{4(1+\nu)}+\ldots\Big)\Big],~ \nu\notin\mathbb{Z},\\
	K_0(x)=&\Big(-\gamma+\frac14(1-\gamma)x^2+\frac1{128}(3-2\gamma)x^4+\ldots\Big)-\log\Big(\frac x2\Big)\Big(1+\frac{x^2}4+\frac{x^4}{64}+\ldots \Big),\\
	K_1(x)=&\frac1x+\frac x4 \Big(2\gamma-\frac18\Big(2\gamma-\frac52\Big)x^2+\ldots\Big)+\frac x2 \log\Big(\frac x2\Big)\Big(1+\frac{x^2}{8}+\ldots\Big),\\
	K_n(x)=&(-1)^{n-1}\log\Big(\frac x2\Big)\sum_{k=0}^{\infty}\frac{(x/2)^{2k}}{k!(k+n)!} +\hf\Big(\frac x2\Big)^{-n}\sum_{k=0}^{n-1}\frac{(-1)^k(n-k-1)!}{k!}\Big(\frac x2\Big)^{2k} \nonumber\\
	&+\frac{(-1)^n}{2}\Big(\frac x2\Big)^n\sum_{k=0}^{\infty}\frac{\psi(k+1)+\psi(k+n+1)}{k!(k+n)!}\Big(\frac x2\Big)^{2k},~n\in\mathbb{N}, \text{ with } \\
	K_{n}(x)=&K_{-n}(x).
\end{align}
Using these, we can write the expansion for $1/f$.
\begin{itemize}
	
	\item For $\nu\notin\mathbb{Z}$ and $\nu<-1$,
	\begin{align*}
		1/f=&\bigg(\hf\Gamma(-\nu)A(p)+\frac{B(p)}{\Gamma(1+\nu)}\bigg)2^{-\nu}z^{D/2+\nu}p^{\nu}\\ &+\bigg(\hf\frac{\Gamma(-\nu)A(p)}{4(1+\nu)}+\frac{B(p)}{\Gamma(2+\nu)}\bigg)2^{2+\nu}z^{D/2+2+\nu}p^{2+\nu} +O(p^{\min{\{-2\nu,4\}}}),
	\end{align*}
	where the order of the rest of the terms is expressed assuming $A(p),B(p)\sim p^{-\nu}$, for which choice, the divergent terms from $1/f$ are removed. 
	
	\item For $\nu=-1$,
	\begin{equation*}
		1/f=\bigg(A(p)+\frac{2B(p)}{\Gamma(0)}\bigg)z^{D/2-1}\frac1p -\hf A(p)z^{D/2+1}p\log{p} +O(p^2),
	\end{equation*}
	taking $A(p),B(p)\sim p$. The $\log$ term makes this nonanalytic.
	
	\item For $-1<\nu<0$,
	\begin{equation*}
	1/f=\bigg(\hf\Gamma(-\nu)A(p)+\frac{B(p)}{\Gamma(1+\nu)}\bigg)2^{-\nu}z^{D/2+\nu}p^{\nu} +2^{-1+\nu}z^{D/2-\nu}\Gamma{(\nu)}A(p)p^{-\nu}+O(p^2),
	\end{equation*}
	again taking $A(p),B(p)\sim p^{-\nu}$, there are no divergent terms. 
	
	\item For $\nu=0$,
	\begin{equation*}
	1/f=-z^{D/2}A(p)\log p -z^{D/2}(A(p)(\gamma+\log{(z/2)})+B(p)) + O(p^2\log p),
	\end{equation*}
	where we must take $A(p), B(p)\sim p^{0}$. This case is impossibly nonanalytic.
	
	\item For $0<\nu<1$,
	\begin{equation*}
			1/f=\Gamma(\nu)A(p)2^{\nu-1}z^{D/2-\nu}p^{-\nu} +\bigg(\hf \Gamma(-\nu)A(p)+\frac{B(p)}{\Gamma(1+\nu)}\bigg)2^{-\nu}z^{D/2+\nu}p^{\nu}+O(p^2),
	\end{equation*}
	with $A(p),B(p)\sim p^{\nu}$. 
	
	\item For $\nu=1$,
	\begin{equation*}
		1/f=A(p)z^{D/2-1}\frac1p -\hf A(p)z^{D/2+1}p\log{p} +O(p^2),
	\end{equation*}
	and we take $A(p)\sim p$ to make the first term analytic. 
	
	\item For $\nu\notin\mathbb Z$ and $\nu>1$,
	\begin{equation*}
		1/f=\Gamma(\nu)A(p)2^{\nu-1}z^{D/2-\nu}p^{-\nu} +\frac{\Gamma(\nu)A(p)}{1+\nu}2^{\nu-1}z^{D/2+2-\nu}p^{2-\nu} +O(p^{\min{\{2\nu,4\}}}),
	\end{equation*}
	with $A(p),B(p)\sim p^\nu$.
\end{itemize}

Thus, we will not be able to do field redefinitions for any integer values of $\nu$, (which is the case for $D=4$). But since there are powers of $p^{2\nu}$ for non-integer values of $\nu$, for half-integer $\nu$, $1/f$ is analytic. This is the case in $D=3$ for the scalar, vector and tensor, where $\nu=\pm1/2,\pm1/2,\pm3/2$, respectively. Other non-integer values of $\nu$ correspond to non-integer dimensions, where we anyway do not expect the Green function behaviour to be analytic.

Now, let us look at the boundary term resulting from redefining the field. From \eqref{boundaryterm}, (see also sec 2.4.3 of \cite{Sathiapalan:2017}), the integrand in the boundary term of the action has the following dependence on $f$.
\begin{equation}
\ddz{\log f}=-f\ddz{\frac1f}.
\end{equation}
The derivatives of the Bessel functions are
\begin{align}
\ddx I_{\nu}(x)=&\frac12(I_{\nu-1}(x)+I_{\nu+1}(x)),\\
\ddx K_{\nu}(x)=&-\frac12(K_{\nu-1}(x)+K_{\nu+1}(x)).
\end{align}
Thus,
\begin{align}
f\ddz \frac{1}{f} &=\frac{D}{2z}-\frac{pA(p)(K_{\nu-1}(pz)+K_{\nu+1}(pz))-pB(p)(I_{\nu-1}(pz)+I_{\nu+1}(pz)} {2(A(p)K_\nu (pz) + B(p)I_\nu(pz))}.
\end{align}

The behaviour of this quantity as $z\to0$ needs to be examined case by case.
\begin{itemize}
	
\item $\pmb{\nu>1, \nu\notin\mathbb{Z}:}$
\begin{equation*}
\lim_{z \to 0}f\ddz\frac1f =\frac{D-2\nu}{2z} - \Big(\frac{1}{2^{2-\nu}(\nu-1)}-\frac18\Big)zp^2 +O(p^{2\nu}).
\end{equation*}
While the leading and subleading terms are analytic, there are powers of the form $p^{2\nu}$. Therefore, the boundary term is analytic for half-integer $\nu$.

\item $\pmb{\nu=1:}$
\begin{equation*}
	\lim_{z \to 0}f\ddz\frac1f =\frac{D-2}{2z} +zp^2\log p +O(p^2).
\end{equation*}
With the log terms, this is nonanalytic. We cannot redefine fields for which $\nu=1$.

\item $\pmb{0<\nu<1:}$
\begin{equation*}
\lim_{z \to 0}f\ddz\frac1f = \frac{D-2\nu}{2z} + \hf \bigg(\frac2z\bigg)^{1-2\nu}\bigg(\frac1{1-\nu}+\frac{2B(p)}{\Gamma(\nu)^2A(p)}\bigg)p^{2\nu}+O(p^2).
\end{equation*}
This is analytic for $\nu=1/2$.

\item $\pmb{\nu=0:}$
\begin{equation*}
\lim_{z \to 0}f\ddz\frac1f =\frac{D}{2z}-\frac1z \bigg(1-\frac{B(p)}{A(p)\Gamma(0)}\bigg)\frac{1}{\log p} + O(p^2),
\end{equation*}
which is not analytic.

\item $\pmb{-1<\nu<0:}$
\begin{equation*}
\lim_{z \to 0}f\ddz\frac1f =\frac{D+2\nu}{2z}-\frac{A(p)\Gamma(1+\nu)}{2A(p)\Gamma(-\nu)+4B(p)/\Gamma(1+\nu)}\bigg(\frac2z\bigg)^{1+2\nu}p^{-2\nu}+O(p^2).
\end{equation*}
This is analytic for $p=-1/2$.

\item $\pmb{\nu=-1:}$
\begin{equation*}
\lim_{z \to 0}f\ddz\frac1f =\frac{D-2}{2z} +\frac{1}{A(p)+2B(p)/\Gamma(0)}zp^2\log p +O(p^2).
\end{equation*}
With the log terms, this is nonanalytic. We cannot redefine fields for which $\nu=-1$.

\item $\pmb{\nu<-1, \nu\notin\mathbb Z:}$
\begin{align*}
\lim_{z \to 0}f\ddz\frac1f =&\frac{D+2\nu}{2z}-\frac z4\frac{A(p)\Big(\frac{\Gamma(-\nu+1)}{4\nu}+\Gamma(-\nu-1)\Big)-2B(p)\Big(\frac1{\Gamma(2+\nu)}+\frac1{\Gamma(1+\nu)}\Big)}{A(p)\Gamma(-\nu)+2B(p)/\Gamma(1+\nu)}p^2\\
&+O(p^{-2\nu}).
\end{align*}
Again, this is analytic for half-integer $\nu$.

\end{itemize}
\vspace{0.1 in}

As described in section \ref{simulmap}, the values of $\nu$ we need for redefining the scalar composite, the vector, and the tensor are $\pm(-D/2+2),\pm(-D/2+1),\pm D/2$ respectively. $\therefore$, for $D=3$, we can redefine all three fields without a problem, but for $D=4$, the operators requires $\nu=0,\pm1$, and we will not be able to redefine it. But, as the quartic theory itself is not defined in $D=4$, we need not worry about that case.

%
%
%
%
%



%% file: tensorAppendix_input.tex
\section{ERG equation for Tensor composite} \label{tensorAppendix}
The details of the derivation of the ERG equation in the tensor case dealt with in Section \ref{tensor} are given below.

We start with:
\be 
e^{-S_\lm[\si_{\mu\nu},\phi_l^I]}=\int \CD \chi _{\mu\nu}  e^{i\int _x \chi_{\mu\nu}\si^{\mu\nu} }\int \CD \phi_h^I e^{-\hf \int _p  \phi_h^I(p) \DD_h^{-1}(p^2) \phi_h^I(-p)-i\int_p \chi_{\mu\nu}(p)\Theta^{\mu\nu}[\phi_l^I+\phi_h^I](-p)}.
\ee
Note that with $\chi_{\mu\nu}$ being transverse and traceless, the terms $\int_x \chi_{\mu\nu}t^{\mu\nu}[\phi^2]$  and $\int_x\chi_{\mu\nu}(x)\dd^{\mu\nu}\phi \Box \phi(x)$ drop out. Therefore, we obtain:
\be
\int_x \chi_{\mu\nu}(x)\Theta^{\mu\nu}[\phi_l^I+\phi_h^I]=-\int_x \chi_{\mu\nu}(x)\phi_l^I \p^\mu\p^\nu \phi_l^I-\int_x \chi_{\mu\nu}(x)\phi_h^I \p^\mu\p^\nu\phi_h^I -\int_x \chi_{\mu\nu}(x) (2\phi_h^I\p^\mu\p^\nu \phi_l^I).
\ee

\paragraph{ Quadratic in $\phi_h$}
Let us collect the terms in the action that are  quadratic in $\phi_h$:
\be
\int_x\int_y  \{\phi_h^I(x) \hf\underbrace{[ \DD_h^{-1}(x,y) - 2i \chi_{\mu\nu}(x)\p_x^\mu\p_x^\nu\dd(x-y)]}_{O^{-1}(x,y)}\phi_h^I(y).
\ee

\paragraph{ Linear in $\phi_h$} 
\be
-2 i\int_x\int_y \phi_l^I(x)\chi_{\mu\nu}\p_x^\mu\p_x^\nu\dd(x-y) \phi_h^I(y)\equiv -i\int _y j^I(y)\phi_h^I(y),
\ee
where
\be
j_l(y)=2\int_x \phi_l(x)\chi_{\mu\nu}(x)\p^\mu\p^\nu \dd(x-y)=2\chi_{\mu\nu}(y)\p^\mu\p^\nu\phi_l (y ). 
\ee
We have used the transverality of $\chi_{\mu\nu}$, ($\p^\mu\chi_{\mu\nu}=0$), in the last equation.

\vspace{0.1 in}
Thus,
\be
\Psi\equiv e^{-S_\lm[\si_{\mu\nu},\phi_l^I]}= \int \CD \chi_{\mu\nu} e^{i\int \chi_{\mu\nu}\si^{\mu\nu}+i\int \chi_{\mu\nu}\phi_l^I\p^\mu\p^\nu\phi_l^I}\int \CD \phi_h^I e^{-\hf \int _x\int_y \phi_h^I O^{-1}(x,y)\phi_h^I+i\int j^I\phi_h^I}
\ee
\be
=\int \CD \chi_{\mu\nu} e^{i\int \chi_{\mu\nu}\si^{\mu\nu}+i\int \chi_{\mu\nu}\phi_l^I\p^\mu\p^\nu\phi_l^I}
Det^{-\frac{N}{2}}[O(x,y)]e^{-\hf \int_x\int_y j^I(x)O(x,y)j^I(y)}.
\ee
$\Psi$ obeys Polchisnki's equation. This is easy to see.
Before integrating over $\chi_{\mu\nu},\si^{\mu\nu}$, one can think of these fields as fixed background fields interacting with the $\phi^I$ fields. 
Thus, consider
\[
\bar \Psi = \int \CD \phi_h^I e^{-\hf \int _p  \phi_h^I(p) \DD_h^{-1}(p^2) \phi_h^I(-p)-i\int_p \chi_{\mu\nu}(p)\Theta^{\mu\nu}[\phi_l^I+\phi_h^I](-p)}.
\]
 $\bar \Psi$ obeys
 \[
 \frac{\p \bar \Psi}{\p t} = \hf\int_{x,y} \dot \DD_h(x-y)\frac{\dd^2\bar \Psi}{\dd \phi_l^I(x)\dd\phi_l^I(y)}.
 \]
Multiply both sides by $e^{\int \chi_{\mu\nu} \sigma^{\mu\nu} }$, and then integrate over $\chi_{\mu\nu}$. One obtains
\be
 \frac{\p  \Psi}{\p t} = \hf\int_{x,y} \dot \DD_h(x-y)\frac{\dd^2 \Psi}{\dd \phi_l^I(x)\dd\phi_l^I(y)}.
 \ee

We evaluate the functional derivatives, and obtain the time evolution of $S_\lm[\si^{\mu\nu},\phi_l^I]$. In particular, we can set $\phi_l^I=0$, and obtain the evolution of $S_\lm[\si^{\mu\nu},\phi_l^I=0]\equiv S_\lm[\si^{\mu\nu}]$. This is a theory with an IR cutoff $\lm$. The infinite volume physics is recovered when $\lm \to 0$. In practice, we obtain a functional differential equation for $S_\lm[\si^{\mu\nu}]$ that can be solved in perturbation theory.

Let us separate out the terms in $ \Psi$ that depend on $\phi_l^I$.
\[
\Psi'= e^{i\int_u \chi_{\mu\nu} (u)\phi_l^I(u) \p^\mu\p^\nu \phi_l^I(u) -\hf \int_u\int_v j^I(u)O(u,v)j^I(v)}.
\]
\[
\frac{\dd \Psi'}{\dd \phi_l^I(x)}=
[2i\chi_{\mu\nu}(x) \p^\mu\p^\nu \phi_l^I(x) -2\int_u\int_v \chi_{\mu\nu}(u)\p_u^\mu\p_u^\nu\dd(u-x)O(u,v)j^I(v)]\Psi';
\]
\[
\frac{\dd^2 \Psi'}{\dd \phi_l^I(x)\dd \phi_l^I(y)}\bigg|_{\phi_l=0}
= [2i \chi_{\mu\nu}(x)\p_x^\mu\p_x^\nu \dd(x-y) -
4\int_u\int_v \chi_{\mu\nu}(u)\p_u^\mu\p_u^\nu\dd(u-x)O(u,v)\chi_{\rho\si}(v)\p_v^\rho\p_v^\si
\dd(v-y)]\Psi'.
\]
We have dropped terms that vanish when $\phi_l^I=0$. 
\be 
\frac{\dd^2 \Psi'}{\dd \phi_l^I(x)\dd \phi_l^I(y)}
=
[2i \chi_{\mu\nu}(x)\p_x^\mu\p_x^\nu \dd(x-y) -
4\chi_{\mu\nu}(x)\p_x^\mu\p_x^\nu \p_y^\rho\p_y^\si O(x,v)\chi_{\rho\si}(y)
]\Psi',
\ee 
from which we obtain
\[
\int_x\int_y \hf \dot \DD_h(x-y)\frac{\dd^2 \Psi}{\dd \phi_l^I(x)\dd \phi_l^I(y)}|_{\phi_l=0}=\int \CD \chi_{\mu\nu}e^{i\int \chi_{\mu\nu}\si^{\mu\nu}}\times
\]
\be
\int_x\int_y \hf \dot \DD_h(x-y)[2i \chi_{\mu\nu}(x)\p_x^\mu\p_x^\nu \dd(x-y) -
4\chi_{\mu\nu}(x)\p_x^\mu\p_x^\nu \p_y^\rho\p_y^\si O(x,v)\chi_{\rho\si}(y)
]Det^{-\frac{N}{2}}[O(x,y)]\bigg|_{\phi_l=0}.
\ee
The first term in the RH above is of the form
\[
\hf \p_x^\mu \p_x^\nu \dot \DD_h(x-y)|_{x=y}\int _x \chi_{\mu\nu}(x)\approx \dd^{\mu\nu}\int_x\chi_{\mu\nu}(x)\approx 0
\]
because $\chi^\mu_\mu=0$.

In the second term, $O(x,y)=\DD_h(x-y) +O(\chi_{\mu\nu})$. 
The leading term is thus
\[
\int \CD \chi_{\mu\nu}e^{i\int \chi_{\mu\nu}\si^{\mu\nu}}\int_x\int_y 2 \dot \DD_h(x-y)\chi_{\mu\nu}(x)\p_x^\mu\p_x^\nu \p_y^\rho\p_y^\si\DD_h(x-y)\chi_{\rho\si}(y)Det^{-\frac{N}{2}}[O(x,y)]
\]
\[
=-\int_x\int_y 2 \dot \DD_h(x-y)\chi_{\mu\nu}(x)\p_x^\mu\p_x^\nu \p_y^\rho\p_y^\si\DD_h(x-y)\frac{\dd^2}{\dd \si^{\mu\nu}(x)\dd \si^{\rho\si}(y)}\int \CD \chi_{\mu\nu}e^{i\int \chi_{\mu\nu}\si^{\mu\nu}}Det^{-\frac{N}{2}}[O(x,y)].
\]
At lowest order, we get
\begin{equation}
\frac{\p}{\p t}e^{-S_\lm[\si_{\mu\nu}]}=
-\hf \int_x\int_y 4 \dot \DD_h(x-y)\p_x^\mu\p_x^\nu \p_y^\rho\p_y^\si\DD_h(x-y)\frac{\dd^2}{\dd \si^{\mu\nu}(x)\dd \si^{\rho\si}(y)}e^{-S_\lm[\si^{\mu\nu}]}, 
\end{equation}
from which we get \eqref{momergtensor}.

%% file: tensor-compressed_input.tex
\section{$\Theta_{\mu\nu}$ Two Point Function}\label{thetamunu}

We give the two point function of the energy momentum tensor in free field theory. 

\subsection{Position Space}

The improved traceless, (traceless on-shell), EM tensor is
\be	\label{1.1.5}
\Theta_{\mu\nu}=\frac{D}{4(D-1)}t_{\mu\nu}[\phi^2] - \phi t_{\mu\nu}\phi + \frac {D-2}{2D}\dd_{\mu\nu}\phi\Box \phi.
\ee
Define
\be
\langle \phi(x)\phi(y)\rangle = G(x-y);
\ee
\be
G(X) = \frac{c}{(X^2)^{\Dt-1}},
\ee
where $X^\mu=(x-y)^\mu$, and $c=\frac{\Gamma(\Dt)}{2\pi^\Dt (D-2)} = \frac{1}{S^{D-1}(D-2)}$. $S^{n}$ denotes the volume of the unit $n$-sphere.\\
Note that $\Box G(X)= \dd^D(X)$.

The result is:
\begin{align}
	\langle \Theta_{\mu\nu}(x)\Theta_{\rho\si}(y)\rangle =&
	c^2\frac{(D-2)^2}{4(D-1)}\bigg\{ -2 \dd_{\mu\nu}\dd_{\rho\si}+ D[\dd_{\mu\si}\dd_{\nu\rho}+\dd_{\mu\rho}\dd_{\nu\si}]\nonumber\\
	&-2D\frac{ [\dd_{\mu\rho}X_\nu X_\si +\dd_{\mu\si}X_\nu X_\rho+\dd_{\nu\rho}X_\mu X_\si+\dd_{\nu\si}X_\mu X_\rho]}{X^2}\nonumber\\ 
	&+8 \frac{X_\mu X_\nu X_\rho X_\si}{(X^2)^2}\bigg\}\times (X^2)^{-D}.
\end{align}
It can be checked that it is traceless (on shell) in $\mu,\nu$ and $\rho, \si$, and also transverse in each index.

\subsection{Momentum Space}

\begin{align}
	\langle \Theta_{\mu\nu}(q) ~~\Theta_{\rho\si}(-q)\rangle=&
	\frac{2}{D(D-1)}\dd_{\mu\nu}\dd_{\rho\si}(q^2)^\Dt- \frac{1}{D}[\dd_{\mu\rho}\dd_{\nu\si}+\dd_{\mu\si}\dd_{\nu\rho}](q^2)^\Dt\nonumber\\
	 &-\frac{2(D-2)}{D(D-1)}q_\mu q_\nu q_\rho q_\si (q^2)^{\Dt-2}
	-\frac{2}{D(D-1)}[q_\mu q_\nu \dd_{\rho\si}+q_\rho q_\si \dd_{\mu\nu}](q^2)^{\Dt-1}\nonumber\\
	&+\frac{1}{D}[\dd_{\mu\rho} q_\nu q_\si + 3~perm](q^2)^{\Dt-1}\times \gamma,
\end{align}
where
$\gamma$ is
\[
\gamma = \frac{\Gamma(1-\Dt)}{\fpi} \frac{\Gamma(\Dt+1)^2}{\Gamma(D+2)}.
\]

We have already seen that the expression is traceless. One can also check explicitly that it is transverse on each index. It is also possible to write it in a manifestly transverse form:
\begin{align}
	\langle \Theta_{\mu\nu}(q) ~~\Theta_{\rho\si}(-q)\rangle=& \gamma\Big\{\frac{2}{D(D-1)}[(\dd_{\mu\nu}q^2-q_\mu q_\nu)
	(\dd_{\rho\si}q^2-q_\rho q_\si)]\nonumber\\
	&-\frac{1}{D}[(\dd_{\mu\rho}q^2-q_\mu q_\rho)(\dd_{\nu\si}q^2-q_\nu q_\si)]\nonumber\\
	&-\frac{1}{D}[(\dd_{\mu\si}q^2-q_\mu q_\si)(\dd_{\nu\rho}q^2-q_\nu q_\rho)]\Big\}.
\end{align}
